\documentclass{aastex6}
\usepackage{subfigure}
\usepackage{rotating}
\usepackage[perpage,symbol*]{footmisc}
\usepackage{multirow}

\accepted{by ApJ on February 7, 2018}

\shortauthors{Zhen Yan et al.}

\begin{document}

\title{Simultaneous 13~cm/3~cm single-pulse observations of PSR~B0329+54}

\author{Zhen~Yan\altaffilmark{1,8}\footnote{yanzhen@shao.ac.cn},
  Zhi-Qiang~Shen\altaffilmark{1,8},
  R.~N.~Manchester\altaffilmark{2}, C.-Y,~Ng\altaffilmark{3}, P.~Weltevrede\altaffilmark{4},
  Hong-Guang~Wang\altaffilmark{5},
  Xin-Ji~Wu\altaffilmark{6},
  Jian-Ping~Yuan\altaffilmark{7,8},
  Ya-Jun~Wu\altaffilmark{1,8},
  Rong-Bing~Zhao\altaffilmark{1,8},
  Qing-Hui~Liu\altaffilmark{1,8},
  Ru-Shuang~Zhao\altaffilmark{1,9},
  Jie~Liu\altaffilmark{1,9}}

\altaffiltext{1}{Shanghai Astronomical Observatory, Chinese Academy of Sciences, Shanghai 200030, China}
\altaffiltext{2}{CSIRO Astronomy and Space Science, PO Box 76, Epping NSW 1710, Australia}
\altaffiltext{3}{Department of Physics, The University of Hong Kong, Pokfulam Road, Hong Kong, China}
\altaffiltext{4}{Jodrell Bank Centre for Astrophysics, School of Physics and Astronomy, University of
  Manchester, Manchester M13 9PL, UK }
\altaffiltext{5}{School of Physics and Electronic Engineering, Guangzhou University, 510006 Guangzhou, China}
\altaffiltext{6}{Department of Astronomy, Peking University, Beijing 100871, China}
\altaffiltext{7}{Xinjiang Astronomical Observatory, Chinese Academy of Sciences, Urumqi 830011, China}
\altaffiltext{8}{Key Laboratory of Radio Astronomy, Chinese Academy of Sciences, China}
\altaffiltext{9}{University of Chinese Academy of Sciences, Beijing 100049, China}

\begin{abstract}
We have investigated the mode-changing properties of PSR~B0329+54
using 31 epochs of simultaneous 13~cm/3~cm single-pulse observations
obtained with Shanghai Tian Ma 65~m telescope. The pulsar was found in
the abnormal emission mode 17 times, accounting for $\sim$13\% of the
41.6 hours total observation time. Single pulse analyses indicate that
mode changes took place simultaneously at 13~cm/3~cm within a few
rotational periods. We detected occasional bright and narrow pulses
whose peak flux densities were 10 times higher than that of the
integrated profile in both bands. At 3~cm, about 0.66\% and 0.27\% of
single pulses were bright in the normal mode and abnormal mode
respectively, but at 13~cm the occurrence rate was only about
0.007\%. We divided the pulsar radiation window into three components
(C1, C2 and C3) corresponding to the main peaks of the integrated profile.
The bright pulses preferentially occurred at pulse
phases corresponding to the peaks of C2 and C3. Fluctuation spectra
showed that C2 had excess red noise in the normal mode, but broad
quasi-periodic features with central frequencies around 0.12
cycles/period in the abnormal mode. At 3~cm, C3 had a stronger
quasi-periodic modulation centered around 0.06 cycles/period in the
abnormal mode. Although there were some asymmetries in the
two-dimensional fluctuation spectra, we found no clear evidence for
systematic subpulse drifting. Consistent with previous low-frequency
observations, we found a very low nulling probability for B0329+54
with upper limits of 0.13\% and 1.68\% at 13~cm/3~cm respectively.
\end{abstract}


\section{Introduction}
Pulsars are fast rotating, highly magnetized neutron stars that
produce lighthouse-like beams of radio emission from their magnetic
poles. The pulsed radiation can only be observed when the emission
beam is pointing toward the Earth, forming a pulse train with an
extremely stable period. In practice, observers can only detect
individual pulses from comparatively strong pulsars, as pulsars are
intrinsically weak radio sources.

In order to improve the signal-to-noise ratio (S/N) of pulsar
observations, the integrated profile is usually used. The integrated
profile is obtained by summing up a number of individual pulses
synchronously with the rotational period. For most pulsars, the shape
of individual pulses varies dramatically from pulse to pulse, while
the integrated profile is very stable at any particular observing
frequency \citep{hmt75}.  However, integrated profiles for some
pulsars show sporadic changes between two (sometimes more than two)
quasi-stable states; this is called mode changing or switching. Ever
since the first detection of mode changes in PSR~B1237+25
\citep{bac70a}, this phenomenon has been seen in a few dozens of
pulsars in subsequent years \citep{lyn71,mgb81,fbw+81,wf81} with
continued increase in recent times \citep[e.g.,][]{wmj07,bjb+12}.
Most of these pulsars have a complex integrated profile with multiple
components.

Some of the pulsars that show mode changes also exhibit drifting
subpulses \citep{lkr+02,jv04,rwr05}. But the connection between
these two phenomena is not well understood yet. It appears
that young pulsars have the most disordered subpulses and that
drifting subpulses are more often seen in older pulsars \citep{ran86}.
Simultaneous multi-frequency single-pulse observations are the most direct
way to study whether the mode changing and subpulse drifting are wide
band phenomena and their intrinsic relationships. Most previous
observations indicated that these two phenomena took place
synchronously over a wide band. However, there are exceptions, e.g.,
PSR~B0031$-$07, where in simultaneous 328~MHz and 4.85~GHz observations
\citep{smk+05}, only mode A occurred at the same time at both
frequencies while modes B and C only occurred at 328~MHz.

PSR~B0329+54 is a pulsar showing a complex integrated profile which
can be fitted with five Gaussian components \citep{kra94}. Long-term
low-frequency observations indicated that this pulsar shows relatively
frequent mode changes, with the pulsar in an ``abnormal'' mode $\sim$15\% of
the time \citep{bmsh82,cww+11}.

\begin{table}[!hbp]
\centering
\begin{tabular}{c c c c}
\hline
\hline
Parameters    & Values     &   Unit                & Reference \\
\hline
$P$   & 0.714520         &   s                   &   \cite{hlk+04}\\
$\dot P$   &  $2.05\times 10^{-15}$  &    -                 &   \cite{hlk+04} \\
DM          &  26.7641  & cm$^{-3}$~pc     &   \cite{hsh+12}\\
$S_{\rm 1400}$&  203       &  mJy                  &   \cite{lylg95}\\
$\alpha$      &  -1.6      &  -                    &   \cite{lylg95}\\
$D$          &  1.0       &  kpc                  &   \cite{ymw17} \\
$\tau$         &   $5.53\times 10^6$ & yr                   & \\
$B_{s}$    &  $1.22\times 10^{12}$  & G                        &  \\
$B_{LC}$   & 31.5   & G                        & \\
$\dot{E}$     & $2.22\times 10^{32}$   &erg~s$^{-1}$             & \\
\hline
\end{tabular}
\caption{Measured and derived parameters of PSR~B0329+54. Parameters listed are the barycentric period ($P$),
time derivative the barycentric period ($\dot P$), dispersion measure (DM),  mean flux density at 1400 MHz ($S_{\rm 1400}$),
spectral index ($\alpha$), DM-derived distance based on the YMW16 model ($D$), spin-down age ($\tau$), surface magnetic flux
density ($B_{\rm s}$), magnetic field at light cylinder ($B_{\rm LC}$) and spin-down luminosity ($\dot{E}$).}
\label{tab:parameters}
\end{table}

In Table~\ref{tab:parameters}, we list the basic measured and
  derived parameters of PSR~B0329+54.\footnote{Data from the ATNF
    Pulsar Catalogue,
    http://www.atnf.csiro.au/research/pulsar/psrcat \citep{mhth05}.}
At a frequency $\nu$ of 1.4~GHz,
the mean flux density of PSR~B0329+54 is about 230~mJy with a spectral index $\alpha$
($S_{\rm \nu}\propto\nu^\alpha$) about $-1.6$ \citep{lylg95}. Because of
its strong flux density and somewhat flat spectrum, the mode changes
of this pulsar have also been detected at high frequencies, such
as 10.5~GHz \citep{xsg+95}, 14.8~GHz \citep{bs78}, 32~GHz
\citep{kxj+96} and 43~GHz \citep{kjdw97}. Dual-band single-pulse
observations of PSR~B0329+54 have been carried out, for example, at
327~MHz and 2695~MHz \citep{bs78}, 1.4~GHz and 9.0~GHz
\citep{bmsh82}. Except for the 327~MHz/2695~MHz observations which were
carried out with a dual-band receiver, the others made uses of
simultaneous observations on different telescopes. In all of these
observations, the mode changes occurred synchronously at the two
different frequencies.

In this paper, we report simultaneous single-pulse observations of
PSR~B0329+54 in the 13~cm and 3~cm bands (corresponding center
frequencies of 2.3~GHz \& 8.6~GHz) with the Shanghai Tian Ma Radio
Telescope (TMRT) in order to study the mode changes and related
phenomena with a comparatively wide frequency coverage. The
observations and data reduction methods are described in
Section~\ref{sec:obsinfo}. Detail about the analyses of mode changes,
bright narrow pulses, subpulse drifting, and pulse nulling are
presented in Sections~\ref{sec:mode}, \ref{sec:brightpulse},
\ref{sec:drift} and \ref{sec:null}, respectively. Finally, discussion
and conclusion will be given in Section~\ref{sec:discuss}.

\section{Observations and data reduction}
\label{sec:obsinfo}
Observations of PSR~B0329+54 were carried out with the TMRT, which is
a fully steerable 65-m diameter Cassegrain antenna with an active
reflector surface. Seven sets of receivers are installed with the
frequency coverage ranging from 1.25~GHz to 50.0~GHz. Taking the
power-law spectrum radiation with the spectral index of -1.6 into
consideration, the 13~cm/3~cm receiver was selected to do simultaneous
dual-frequency observations of PSR~B0329+54. This 13~cm/3~cm receiver
is a cryogenically cooled, dual-polarization and dual-band receiver
with frequency coverage of 2.2 -- 2.4~GHz and 8.2 -- 9.0~GHz in the 13~cm
and 3~cm bands respectively.  Normally, the system equivalent flux
density (SEFD) of the TMRT is about 46~Jy and 48~Jy at 13~cm and 3~cm,
respectively.

The digital backend system (DIBAS), an FPGA-based spectrometer based
upon the design of VEGAS \citep{abb+12} with pulsar modes that provide
much the same capabilities as GUPPI \citep{drd+08}, was used for the
data sampling and recording. DIBAS has three pairs of digitizers (one
for each polarization pair), making it possible to sample and digitize
data from different bands \citep{ysw+15}. Our observations used pulsar
search mode and the signals from the 13~cm and 3~cm were sampled by
the first two digitizer pairs. The received bandwidths, 200~MHz and
800~MHz for 13~cm and 3~cm respectively, were each divided into 512
channels to allow mitigation of radio-frequency interference (RFI) and
off-line dedispersion. In order to reduce the data rate, we chose to
record total intensity (i.e.~Stokes I) only by summing the two
polarization channels after digitization. The data from each receiver
were written out separately as an 8-bit PSRFITS \citep{hvm04}
file. Observing parameters are listed in the first four columns of
Table~\ref{tab:obsinfo}. They are, respectively, the epoch number
($N$), observation start time as a Modified Julian Date (MJD),
observation length ($T_{\rm o}$) and sampling interval ($t_{\rm samp}$). At
the time of our observations, the diodes that inject pulsed signals
into the front-end of each receiver had not been installed. Before
starting each observation, a careful adjustment was made to ensure
that the power levels of the left-hand and right-hand polarization
channels were at the level of $-$20.0~dBm with an accuracy better
than 0.5~dBm.

\begin{table}[!hbp]
\small
\centering
\begin{tabular}{c c c c c c c c}
\hline\hline
$N$ & $\rm MJD$ & $T_{\rm o}~(s)$&$t_{\rm samp}~(\mu s)$ & $M_{\rm A}$ & $T_{\rm A}$~(s)& $S_{\rm A/N,13cm}$ & $S_{\rm A/N,3cm}$ \\
\hline
1 &   56755.33 &   1636 &   65.54  &       Y  & $590\pm5$ & $0.08 \pm 0.01$ & $0.21\pm0.14$ \\
2 &   56759.10 &   4648 &   65.54  &       Y  & $875\pm3$ &$0.47 \pm 0.01 $ & $0.16\pm0.11$\\
3 &   56759.17 &   3729 &   65.54  &       N  & 0  & -& - \\
4 &   56759.23 &   2857 &   65.54  &       N  & 0 &- &-\\
5 &   56759.42 &   2641 &   65.54  &       Y  & $2641\pm1$ &- &- \\
6 &   56761.08 &   1685 &   65.54  &       Y  & $1685\pm1$ \\
7 &   56782.12 &   1866 &   32.77  &       N  & 0 &- &- \\
8 &   56782.19 &   2556 &   32.77  &       N  & 0&- &- \\
9 &   56782.28 &   3399 &   32.77  &       N  & 0 &- &-\\
10 &   56784.07 &   7187 &   65.54  &       N  & 0 &- &- \\
11 &  56784.15 &   7187 &   131.07 &       N  & 0 &- &- \\
12 &  56784.24 &   7187 &   131.07 &       Y  & $782\pm4$ &$-0.55\pm0.01$ & $0.29\pm0.08$\\
13 &  56784.32 &   4233 &   131.07 &       Y  & $875\pm4$ &$0.22\pm 0.02$ & $0.05\pm0.06$\\
14 &  56785.08 &   7188 &   131.07 &       N  & 0  &-& -\\
15 &  56785.12 &   7188 &   131.07 &       Y  & $222\pm5$ &$-0.43\pm 0.02$ & $-0.12\pm0.07$\\
16 &  56785.20 &   7188 &   131.07 &       Y  &$1477\pm6/235\pm3$ & $0.37\pm0.01$& $0.02\pm0.03$ \\
17 &  56785.28 &   7188 &   131.07 &       Y  & $4579\pm3$ &$-0.12\pm0.01$& $0.10\pm0.02$\\
18 &  56786.04 &   7188 &   131.07 &       N  & 0  &- &-\\
19 &  56786.12 &   1626 &   131.07 &       N  & 0 &- &-\\
20 &  56786.31 &   5594 &   131.07 &       Y  & $154\pm2/289\pm3$ & $0.01\pm0.03$ & $0.21\pm0.04$ \\
21 &  56836.04 &   7188 &   131.07 &       N  & 0  &- &-\\
22 &  56836.16 &   6733 &   131.07 &       Y  & $245\pm6$ &$0.30\pm0.02$ & $0.11\pm0.14$\\
23 &  56850.11 &   3580 &   131.07 &       N  & 0 &- &-\\
24 &  56863.06 &   2437 &   131.07 &       N  & 0  &- &- \\
25 &  56863.10 &   7188 &   131.07 &       N  & 0  &- &-  \\
26 &  56863.96 &   4061 &   131.07 &       Y  & $2153\pm7$ &$0.51\pm0.02$& $0.30\pm0.04$  \\
27 &  56950.62 &   5395 &   65.54  &       Y  & $900\pm6$& $0.23\pm0.01$ & $0.23\pm0.03$  \\
28 &  56950.68 &   5395 &   65.54  &       N  & 0  &- &- \\
29 &  56950.74 &   5395 &   65.54  &       N  & 0  &- &- \\
30 &  56950.81 &   6189 &   65.54  &       Y  & $1486\pm4$ & $-0.20\pm0.01$ & $0.09\pm0.02$\\
31 &  56951.68 &   2396 &   65.54  &       Y  & $90\pm5$ &$-0.11\pm0.05$ & $0.21\pm0.14$  \\
\hline
\hline
\end{tabular}
\caption{Observation parameters and mode-change statistics for
  PSR~B0329+54. The epoch number ($N$), observation start time as a
  Modified Julian Date (MJD), observation length ($T_{\rm o}$) and
  sampling interval ($t_{\rm samp}$) are listed in the first four
  columns of this table respectively. The marks `Y' and `N' given in
  the fifth column ($M_{\rm A}$) indicates detection or non-detection
  of the abnormal mode in the corresponding observation. The length of
  the abnormal mode ($T_{\rm A}$) is presented in the sixth column. In
  cases where there were two or more blocks of abnormal mode in that
  epoch of observation, the duration of each abnormal mode block is
  given. The final two columns give the parameters
   $S_{A/N} \equiv (S_{\rm A}-S_{\rm N})/(S_{\rm A}+S_{\rm N})$ for
   the 13~cm band and 3~cm band, respectively, where $S_{\rm A}$ and
   $S_{\rm N}$ are mean pulsed flux densities for the abnormal and
   normal modes correspondingly.}
\label{tab:obsinfo}
\end{table}

At 13~cm, telecommunication RFI seriously affected the observational
data, with only about 100~MHz of reasonably clean bandwidth left for
each epoch of observation. At 3~cm, RFI was seldom a problem and the
entire 800~MHz bandwidth was kept for further analyses. After RFI
rejection, channels were summed using incoherent dedispersion to give
a single data stream for each band. These were converted to
single-pulse time series by folding the data at the predicted
topocentric period using polynomial coefficients generated with TEMPO2
\citep{hem06}. The single-pulse series were written out with 1024
phase bins per period in the PSRFITS format. The PSRCHIVE programs
\citep{hvm04} were used for further data editing and processing.

\section{Mode changes in PSR~B0329+54}\label{sec:mode}
\subsection{Mode change detection}
\begin{figure}[ht]
\centering
\includegraphics[width=0.7\textwidth,angle=-90]{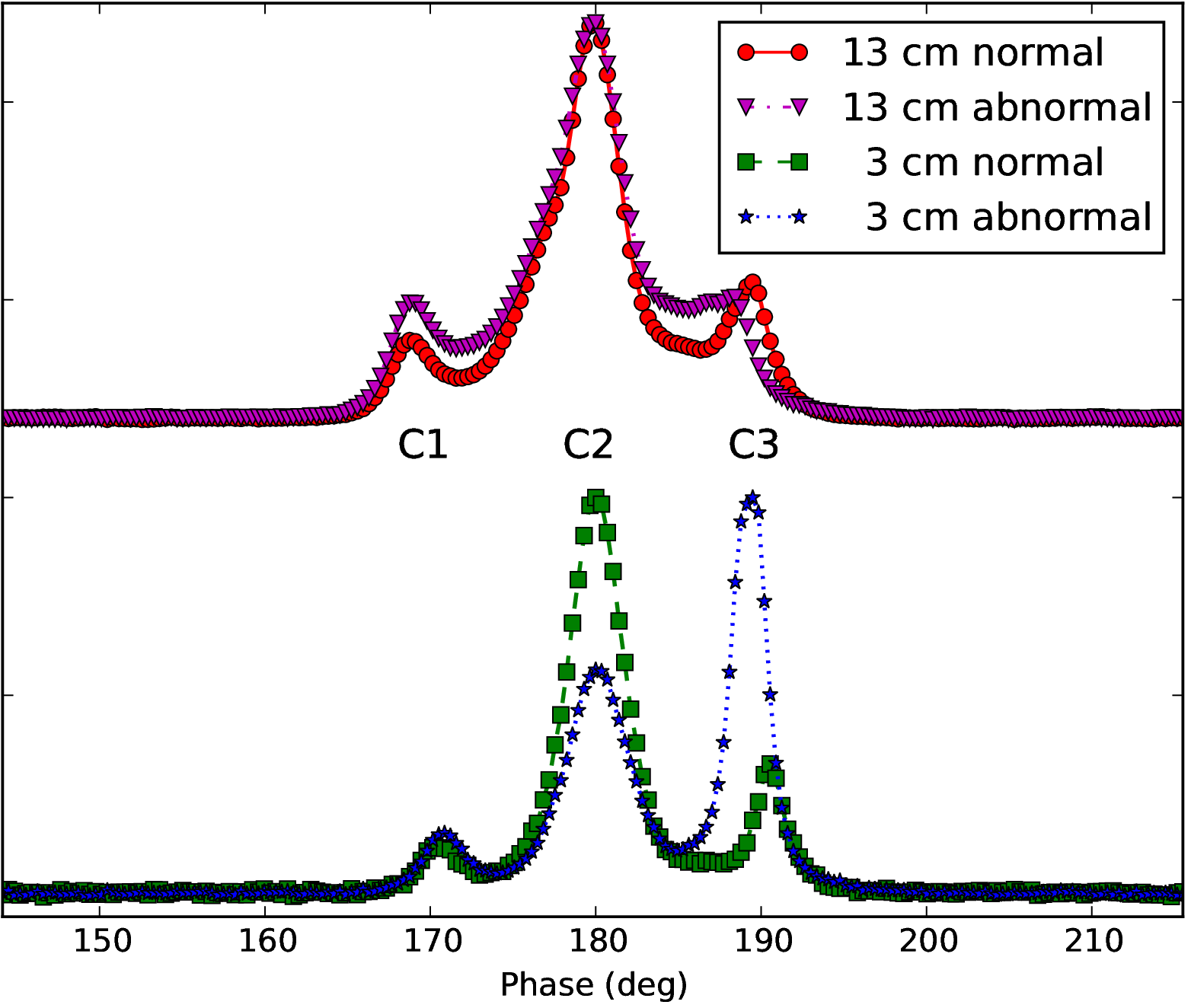}
\caption{Typical normal and abnormal mode integrated profiles of
  PSR~B0329+54 obtained at 13~cm/3~cm with the principal pulse
  components identified.}
\label{fig:inteprof}
\end{figure}

Pulsar mode changes are sporadic and abrupt switches of the integrated
profile between at least two quasi-stable states.  In
Figure~\ref{fig:inteprof}, we present typical normal and abnormal
integrated profiles of PSR~B0329+54 obtained with our 13~cm/3~cm
observations.
At both 13~cm and 3~cm the mean pulse profiles have three main peaks denoted by
C1, C2 and C3. The most significant feature of the
mode change in PSR~B0329+54 is the change in the peak strength and
position of C3, especially at 3~cm. At 13~cm, C3 moves to an
earlier phase, decreases in strength and merges with the bridge
between C2 and C3. At 3~cm, similar to 13~cm, C3 moves to earlier phase (by about
$1\degr$) but, unlike at 13~cm, increases considerably in strength, so
that it is stronger than C2.

\begin{figure}[ht]
\par\vspace{5pt}
\centering
\includegraphics[width=0.78\textwidth,angle=-90]{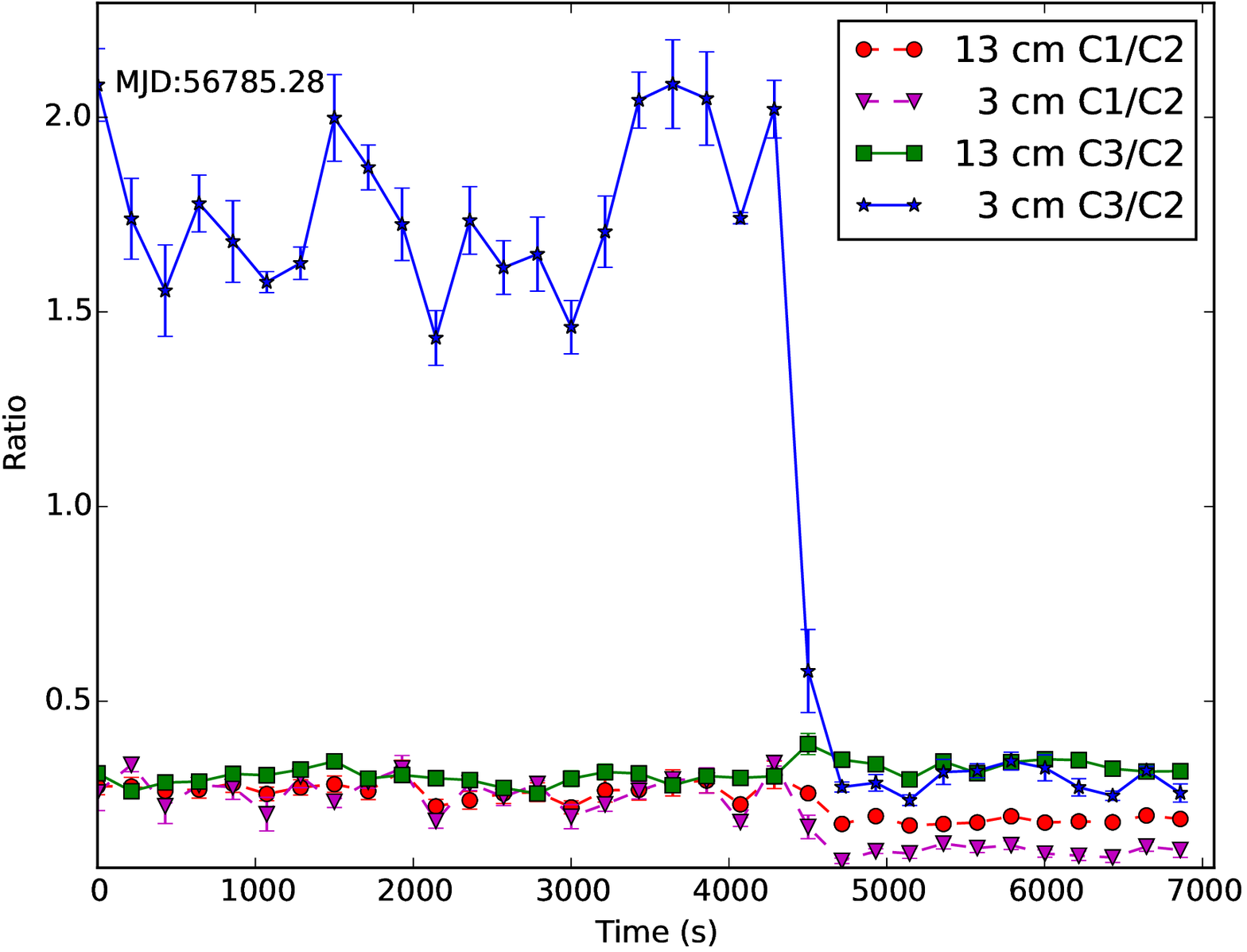}
\caption{Time sequence of peak intensity ratios C1/C2 and C3/C2 at
  13~cm and 3~cm for PSR~B0329+54. This sequence started at
  MJD~56785.28.}
\label{fig:peakratio}
\end{figure}

The peak intensity ratio of the leading (C1) and the trailing (C3)
components relative to the central peak (C2), labeled as C1/C2 and
C3/C2, were used as indicators for potential mode changes in the
PSR~B0329+54 observations. We found that relatively stable integrated
profiles of PSR~B0329+54 can be obtained by averaging about 300
adjacent pulses. We therefore performed a mode change analysis with
the sub-integration time equal to 300 rotational periods.  In
Figure~\ref{fig:peakratio}, we show an example of a time sequence of
the peak intensity ratios C1/C2 and C3/C2. During this
epoch of observation, PSR~B0329+54 changed from the abnormal to the
normal mode. The plot shows that the ratio C1/C2 exhibited similar
variations at both 13~cm and 3~cm, decreasing from 0.28 and 0.26 to 0.19
and 0.12, respectively, during the mode change. Meanwhile, C3/C2 showed
very different behavior at 13~cm and 3~cm; it increased from 0.30 to
0.34 at 13~cm but decreased dramatically from 1.60 to 0.33 at 3~cm.

\begin{sidewaysfigure}
\centering
\begin{minipage}[b]{0.18\textwidth}
\includegraphics[width=\textwidth,angle=0]{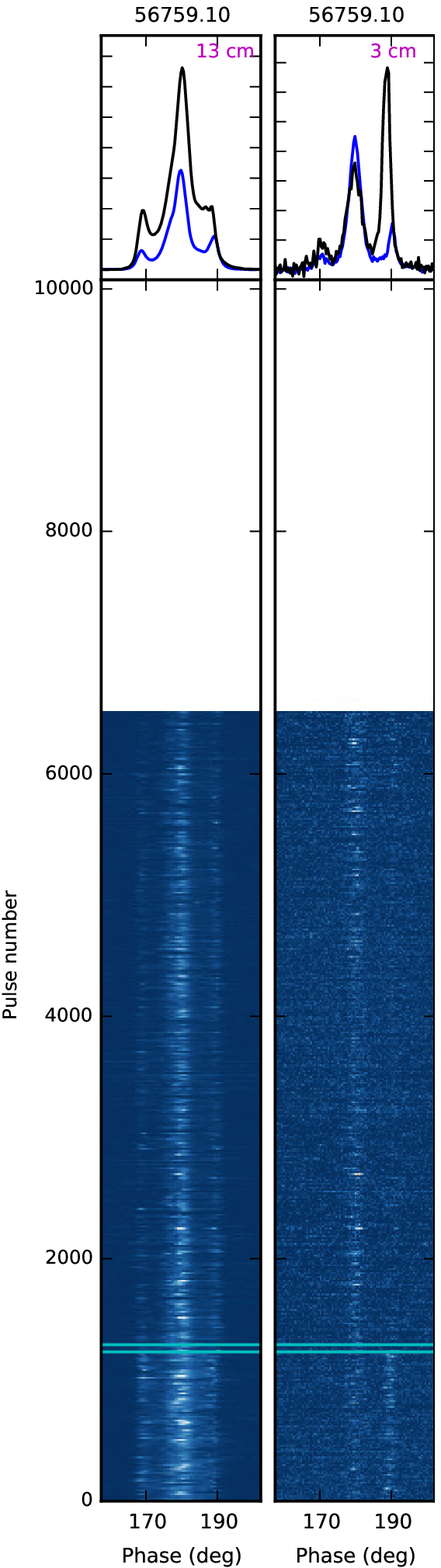}
\end{minipage}
\begin{minipage}[b]{0.18\textwidth}
\includegraphics[width=\textwidth,angle=0]{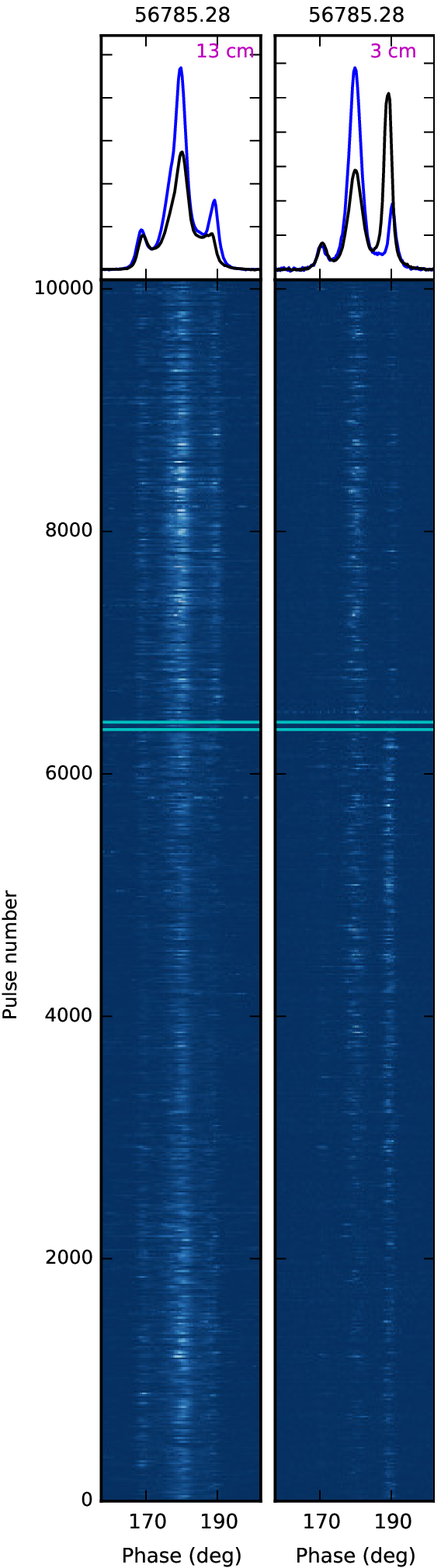}
\end{minipage}
\begin{minipage}[b]{0.18\textwidth}
\includegraphics[width=\textwidth,angle=0]{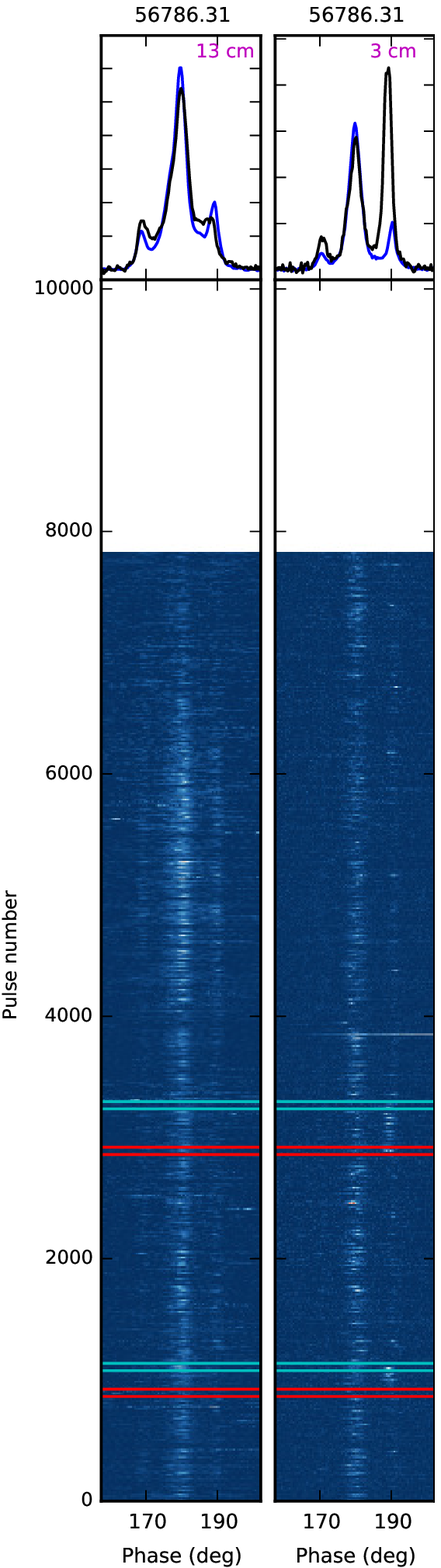}
\end{minipage}
\begin{minipage}[b]{0.18\textwidth}
\includegraphics[width=\textwidth,angle=0]{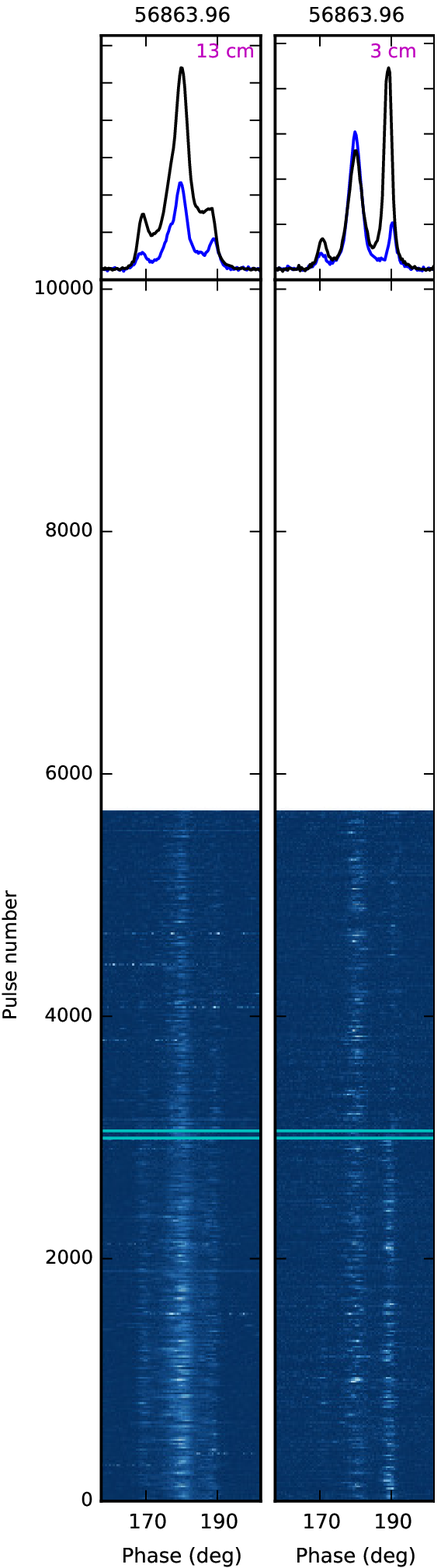}
\end{minipage}
\begin{minipage}[b]{0.18\textwidth}
\includegraphics[width=\textwidth,angle=0]{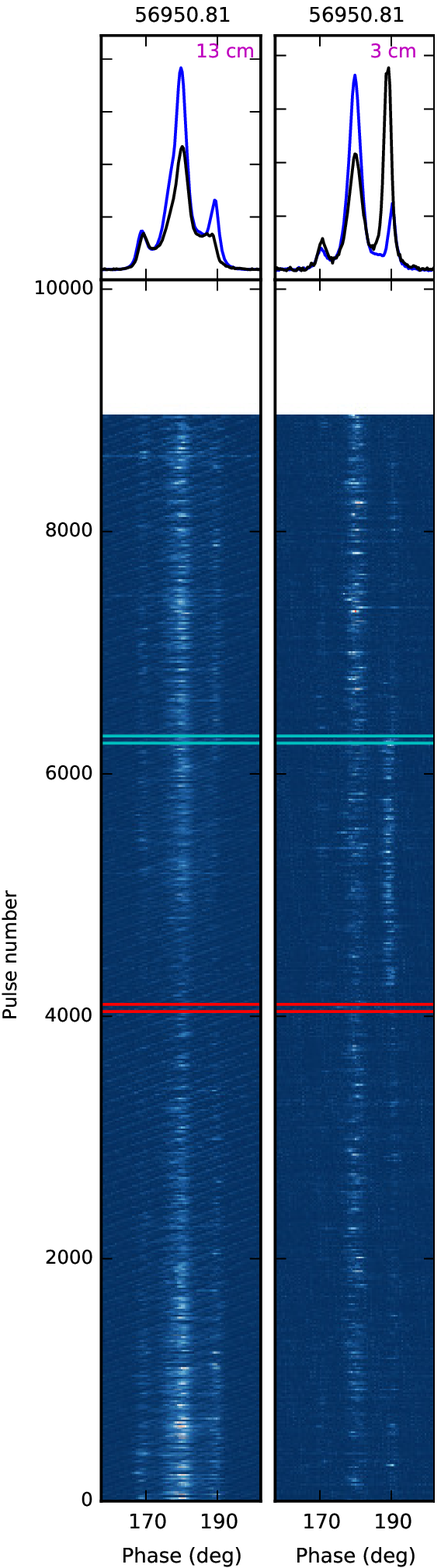}
\end{minipage}
\caption{Samples of 13~cm and 3~cm phase-time plots for mode changes
  of PSR~B0329+54. Top panel: the normal (blue) and abnormal (black)
  mode integrated profiles obtained with all the data of the
  corresponding mode in that epoch of observation.  The MJD and
  observation band are labeled on the top and top right corner,
  respectively. Bottom panel: phase-time plots of single-pulse data
  stacked upwards. The brightness is linear in flux density with
  lighter being stronger. The red line pairs bracket 60 pulses around
  transitions from normal mode to abnormal mode and, similarly, cyan
  line pairs bracket transitions from abnormal mode to normal mode. }
\label{fig:phase-time}
\end{sidewaysfigure}

The peak ratio method discussed above cannot identify mode changes
shorter than about 300 rotation periods. For strong pulsars like PSR
B0329+54, the mode change phenomenon can easily be discerned in the
single-pulse phase-time plots in which the pulse strength is expressed
with a color gradient. In Figure~\ref{fig:phase-time}, we present a
sample of phase-time plots that show mode changes on MJD~56759.10
($N$=2), MJD~56785.28 ($N$=17), MJD~56786.31 ($N$=20), MJD
56863.96~($N$=26) and MJD~56950.81 ($N$=30), along with the
normal and abnormal mode integrated profiles obtained with all the
data of the corresponding mode in that epoch of observation. The time
resolution of the phase-time plots is much higher compared with the
peak ratio plots shown in Figure~\ref{fig:peakratio}.

\begin{sidewaysfigure}
\centering
\subfigure[Normal to abnormal mode]{
\begin{minipage}[b]{0.48\textwidth}
\fbox{\includegraphics[width=0.95\textwidth,angle=0]{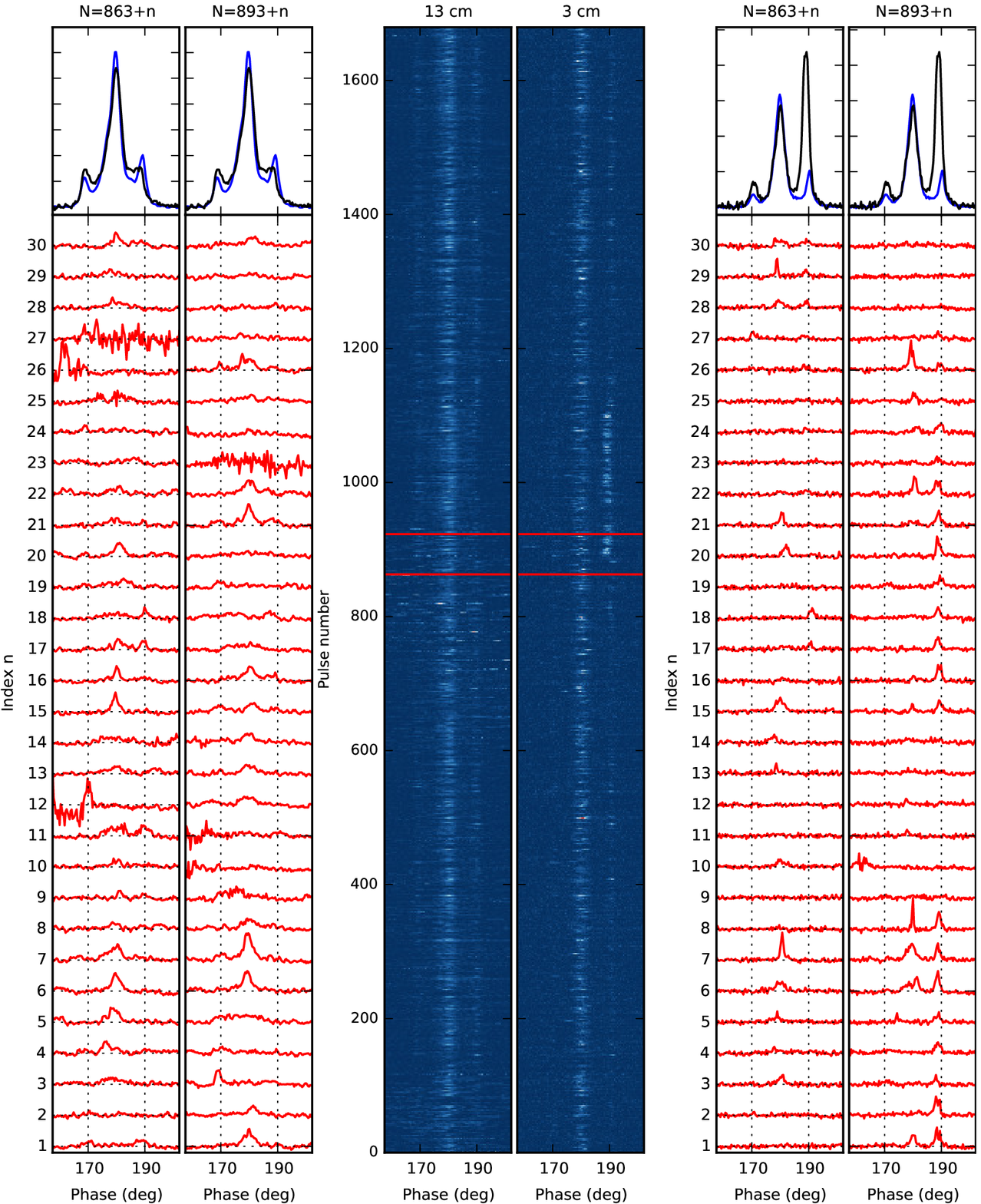}}
\end{minipage}
}
\subfigure[Abnormal to normal mode]{
\begin{minipage}[b]{0.48\textwidth}
\fbox{\includegraphics[width=0.95\textwidth,angle=0]{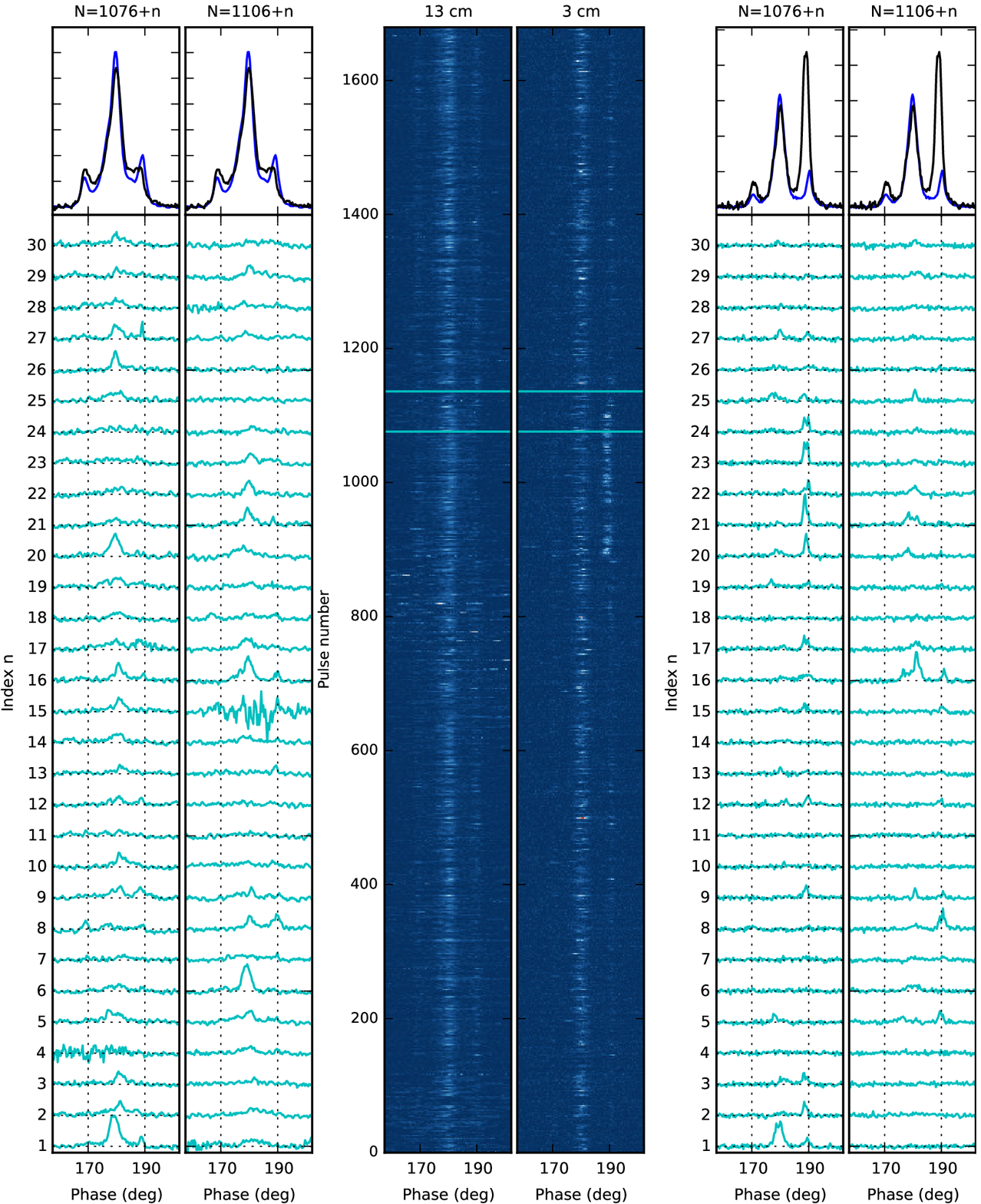}}
\end{minipage}
}
\caption{Single-pulse phase-time plots around PSR~B0329+54 mode
  changes. (a) A normal to abnormal mode change at 13~cm (left panels)
  and 3~cm (right panels). The central colour plots show about 1600
  pulses around the mode change and the left and right panels show
  single-pulse profiles for the 60 pulses centered on the change as
  bracketed by the red lines in the central plots. In both the left
  (13~cm) and right (3~cm) profile plots, the first 30 pulses are
  shown in the left sub-panel and the second 30 in the right sub-panel
  according to the pulse number specifications at the top of each
  sub-plot. The integrated profiles shown at the top of each sub-panel
  are for normal (blue) and abnormal (black) modes, obtained
  with the all the data for the corresponding mode and band in that epoch of
  observation. (b) Similar plots for an abnormal to normal mode
  transition.}
\label{fig:singlepul}
\end{sidewaysfigure}

Figure~\ref{fig:singlepul} shows expanded views of mode changes for both
normal to abnormal (a) and abnormal to normal (b) transitions. Single-pulse
profiles across each transition at both 13~cm and 3~cm are shown. These
and similar plots for other transitions show that the mode transitions
in PSR~B0329+54 occurred abruptly and simultaneously at 13 and 3~cm
within a few pulse rotational periods.

We applied the two methods discussed above to identify mode changes in
the 31 epochs of the TMRT observations (see Table~\ref{tab:obsinfo}).
Except the normal and abnormal pulsar modes
described above, there was no evidence for other mode
states. Table~\ref{tab:obsinfo} shows that the abnormal mode was
identified in 17 of the 31 observations, two separate bursts seen in
two observations ($N$ = 16 and 20). The duration of the abnormal
mode ranges between 1.5 minutes and 1.25 hours. Uncertainties in
the duration were estimated from the uncertainties in identifying the
appearance and disappearance of component C3 in the stacked single
pulse phase-time plots. Summing the abnormal-mode times shows that
PSR~B0329+54 was in the abnormal mode for about 13\% of total
observing time. This is slightly less than $\sim$15\% reported
previously \citep{bmsh82,cww+11}.

\subsection{Quantitative analysis of integrated profiles}
The integrated profiles for PSR B0329+54 in the two modes at the two frequency
bands shown in Figure~\ref{fig:inteprof} are normalised to their peak
values. Here we compare flux density variations
in the abnormal and normal modes at the two bands.
Flux densities were estimated from the signal-to-noise ratio of the data and the
theoretical root-mean-square baseline noise using the corresponding
SEFD of TMRT given in Section~\ref{sec:obsinfo}.
In order to make the flux density changes more easily discernable, the
following quantities were estimated:
\begin{equation}
S_{\rm A/N,13cm}=(S_{\rm A,13cm}-S_{\rm N,13cm})/(S_{\rm A,13cm}+S_{\rm N,13cm}),
\end{equation}
\begin{equation}
S_{\rm A/N,3cm}=(S_{\rm A,3cm}-S_{\rm N,3cm})/(S_{\rm A,3cm}+S_{\rm N,3cm}),
\end{equation}
where $S_{\rm A,13cm}$, $S_{\rm N,13cm}$, $S_{\rm A,3cm}$, $S_{\rm N,3cm}$ are the mean flux
densities at 13~cm and 3~cm in the abnormal and normal modes, respectively.
For each calculation, we only used the data from that epoch of observation. This minimizes the
effects of uncertain calibration, assuming that the instruments were
stable during observations shorter than 2 hours. Derived values of $S_{\rm A/N,13cm}$ and
$S_{\rm A/N,3cm}$ are listed in the last two columns of
Table~\ref{tab:obsinfo}.

It is clear that $S_{\rm A/N,13cm}$ and
$S_{\rm A/N,3cm}$ show different fluctuation properties. The values of
$S_{\rm A/N,13cm}$ are quite variable, but seem to distribute randomly
around 0.0, whereas almost all the $S_{\rm A/N,3cm}$ values are greater
than 0.0 except for that on MJD~56785.12. Quantitatively, the mean
value and rms deviation of $S_{\rm A/N,13cm}$ are $0.06\pm0.32$,
whereas for $S_{\rm A/N,3cm}$ they are $0.14\pm0.11$. It is clear that
the flux density of PSR~B0329+54 at 3~cm is larger in the abnormal
mode than in the normal mode.

Further, we attempt to quantify parameters of the integrated profiles by
fitting five Gaussian components to C1, C2, C3 and the bridge emission, and
then calculating the peak flux density ratios $S_{\rm C1/C2}$ and $S_{\rm C3/C2}$,
and component separations $\Delta\phi_{\rm C1,C2}$ and $\Delta\phi_{\rm C3,C2}$.
Our results indicate that these quantities are stable from observation to
observation within the uncertainties. Table~\ref{tab:intprof} lists
the mean values and their rms deviations across all observations. We
also list values of the overall pulse width at 10\% of the mean profile peak,
$W_{10}$, derived from the Gaussian components.

\begin{table}[!hbp]
\centering
\caption{Parameters for the PSR~B0329+54 integrated profiles for the
  normal and abnormal modes and for the 13~cm and 3~cm bands.}
\begin{tabular}{c c c c c}
\hline
\hline
\multirow{2}{*}{Parameter} & \multicolumn{2}{c}{13~cm}   & \multicolumn{2}{c}{3~cm}   \\
\cline{2-5}
                  & Normal        & Abnormal          & Normal        & Abnormal \\
\hline
$S_{\rm C1/C2}$   & $0.19\pm0.01$ & $0.28\pm0.01$     & $0.11\pm0.01$ & $0.26\pm0.01$ \\
$S_{\rm  C3/C2}$  & $0.34\pm0.01$ & $0.31\pm0.01$     & $0.33\pm0.02$ & $1.63\pm0.05$ \\
$\Delta\phi_{\rm C1,C2}$ (\degr) & $10.67\pm0.06$  & $10.81\pm0.11$ & $9.52\pm0.08$ & $9.27\pm0.13$  \\
$\Delta\phi_{\rm C3,C2}$ (\degr) & $9.36\pm0.09$ & $8.07\pm0.17$ & $10.51\pm0.04$ & $9.50\pm0.10$ \\
$W_{\rm 10}$~(\degr) & $24.05\pm0.18$ & $23.30\pm0.32$ & $22.29\pm0.21$ & $22.55\pm0.28$ \\
$\rho$~(\degr) & $6.8\pm0.3$& $6.6\pm0.3$& $6.4\pm0.3$ & $6.4\pm0.3$ \\
$r$~(km) & $213.3\pm18.3$ &$202.0\pm19.9$ & $187.4\pm17.6$ & $191.1\pm18.8$  \\
\hline
\end{tabular}
\label{tab:intprof}
\end{table}

\section{Narrow bright pulses}
\label{sec:brightpulse}
In the phase-time plots presented in Figure~\ref{fig:phase-time}, some
narrow bright pulses are evident. Similar bright pulses have been
detected in several normal pulsars, including PSRs B0656+14
\citep{wsrw06}, B0943+10 \citep{bmr10}, and B0031$-$07 \citep{kss11},
and a small number of magnetars, including PSRs~J1809$-$1943
\citep{ssw09}, J1622$-$4950 \citep{lbb12}, and J1745$-$2900
\citep{ysw+15}. These pulses are generally five to ten times the
average pulse amplitude, but are not as bright as the ``giant'' pulses
seen in the Crab and other pulsars (e.g., Mickaliger et al.~2012;
Knight et al.~2006 \nocite{mml+12,kbm+06}). True giant
pulses are also distinguished by their power-law amplitude
distributions as compared to typically log-normal distributions for
other pulse emission including the bright narrow pulses.

\begin{figure}[ht]
\centering
\par\vspace{10pt}
\subfigure[Bright pulse in Abnormal mode)]{
\begin{minipage}{0.88\textwidth}
\includegraphics[width=0.2\textwidth,angle=-90]{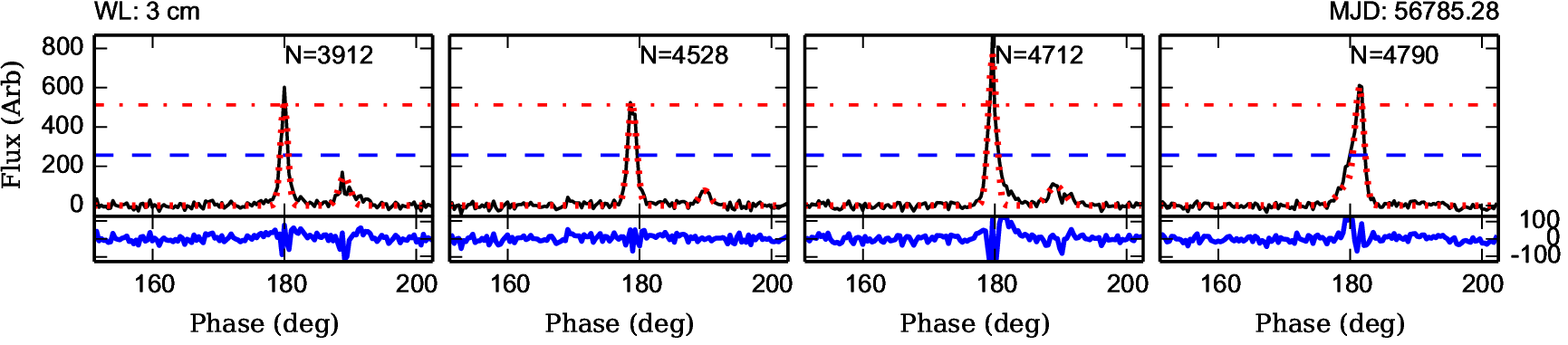}
\end{minipage}
}
\par\vspace{14pt}
\subfigure[Bright pulse in Normal mode]{
\begin{minipage}{0.88\textwidth}
\includegraphics[width=0.2\textwidth,angle=-90]{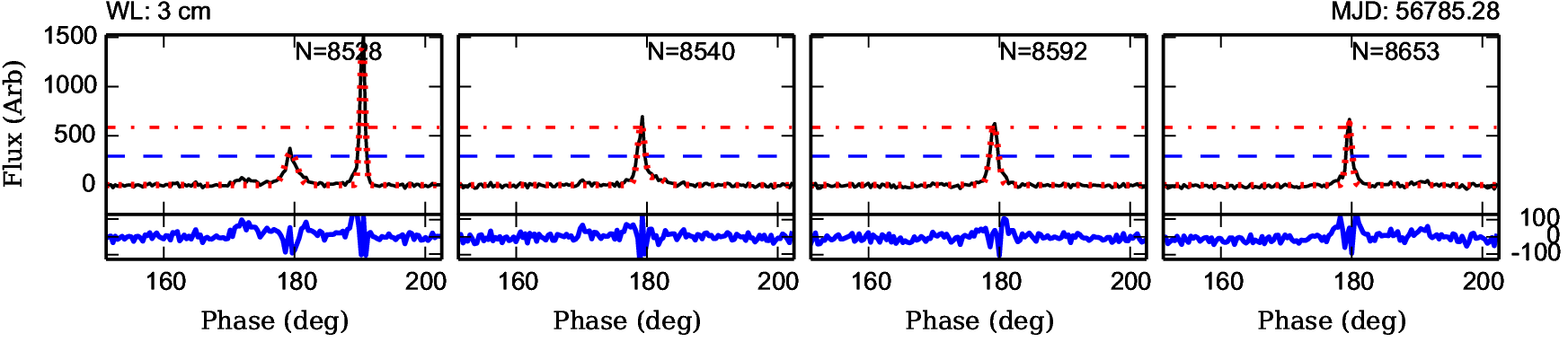}
\end{minipage}
}
\caption{Bright narrow pulses in the 3~cm band from PSR~B0329+54. The dashed and
  dot-dashed lines are at 5 and 10 times the peak flux density of the
  integrated profile ($S_{\rm int,pk}$) respectively. The dotted lines
  show the profiles resulting from the Gaussian fitting. Residuals
  from these fits are shown in the bottom sub-panel of each plot.}
\label{fig:bright}
\end{figure}

Profile plots for a sample of bright pulses from PSR B0329+54 are
shown in Figure~\ref{fig:bright}.  We fitted Gaussian components to
characterize these bright pulses and to determine any quantitative
differences from other pulses. The results of this fitting are shown
in Figure~\ref{fig:bright}. To quantify the strength of the bright pulses,
they are normalized by the peak flux of the corresponding integrated
profile ($S_{\rm int,pk}$). As the integrated profile changes with the
pulsar mode, pulses obtained in the normal and abnormal modes are
treated separately. According to their peak flux density $S_{\rm
  pk}$, pulses were classified by three groups: $<5$~times,
$5-10$~times, and $>10$~times of $S_{\rm int,pk}$. We found that most
of the bright pulses can be fitted with a single Gaussian
component. For the sample in Figure~\ref{fig:bright}, we
chose pulses whose $S_{\rm pk}$ exceeds 10 times of
$S_{\rm int,pk}$ in both the normal and abnormal modes.

Compared with the 3~cm results, at 13~cm there were many fewer bright
pulses with $S_{\rm pk}$ exceeding 10 times of $S_{\rm int,pk}$. For
the 3~cm normal and abnormal modes, the occurrence rates were 0.66\%
and 0.27\% respectively, whereas at 13~cm, the total rate was only
about 0.007\%.

\begin{figure*}[ht]
\centering
\begin{minipage}{0.8\textwidth}
\includegraphics[width=0.7\textwidth,angle=-90]{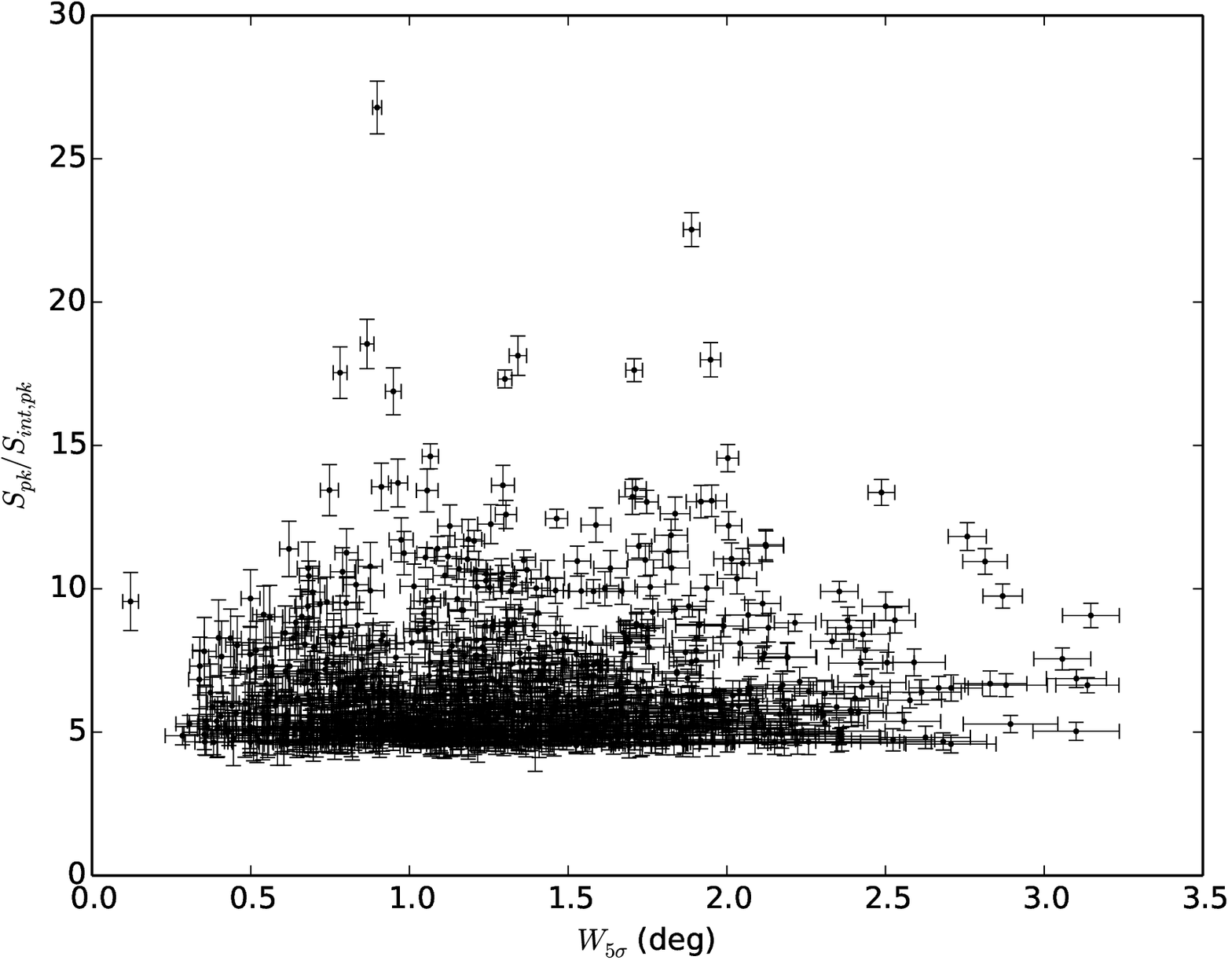}
\end{minipage}
\par\vspace{25pt}
\begin{minipage}{0.8\textwidth}
\includegraphics[width=0.7\textwidth,angle=-90]{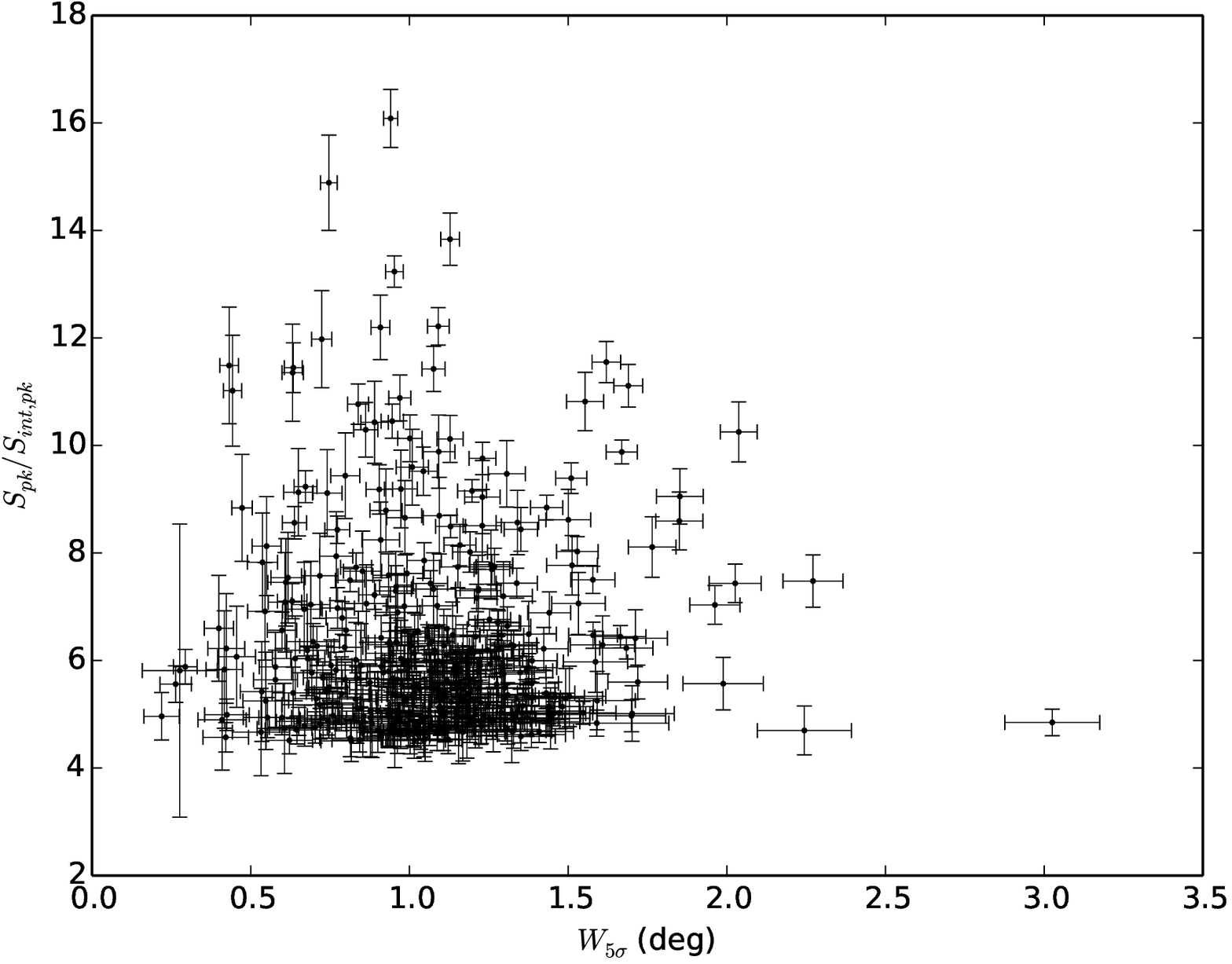}
\end{minipage}
\caption{Scatter plots of $W_{\rm 5\sigma}$ versus $S_{\rm pk}/S_{\rm int,pk}$
for single pulses whose peak flux density exceeds $5S_{\rm int,pk}$
for the normal mode (upper panel)
and the abnormal mode (lower panel). }
\label{fig:widthpeak}
\end{figure*}

Both the width at 5 times the rms baseline noise ($W_{\rm 5\sigma}$) and
$S_{\rm pk}$ of bright single pulses were obtained from the Gaussian fitting.
As Figure~\ref{fig:widthpeak} shows, stronger single pulses tend to be narrower
than average, but there is no clear correlation between the peak flux density
and the width of the pulses.

\begin{figure*}[ht]
\centering
\subfigure[Normal mode]{
\begin{minipage}[b]{0.39\textwidth}
\includegraphics[width=0.88\textwidth,angle=-90]{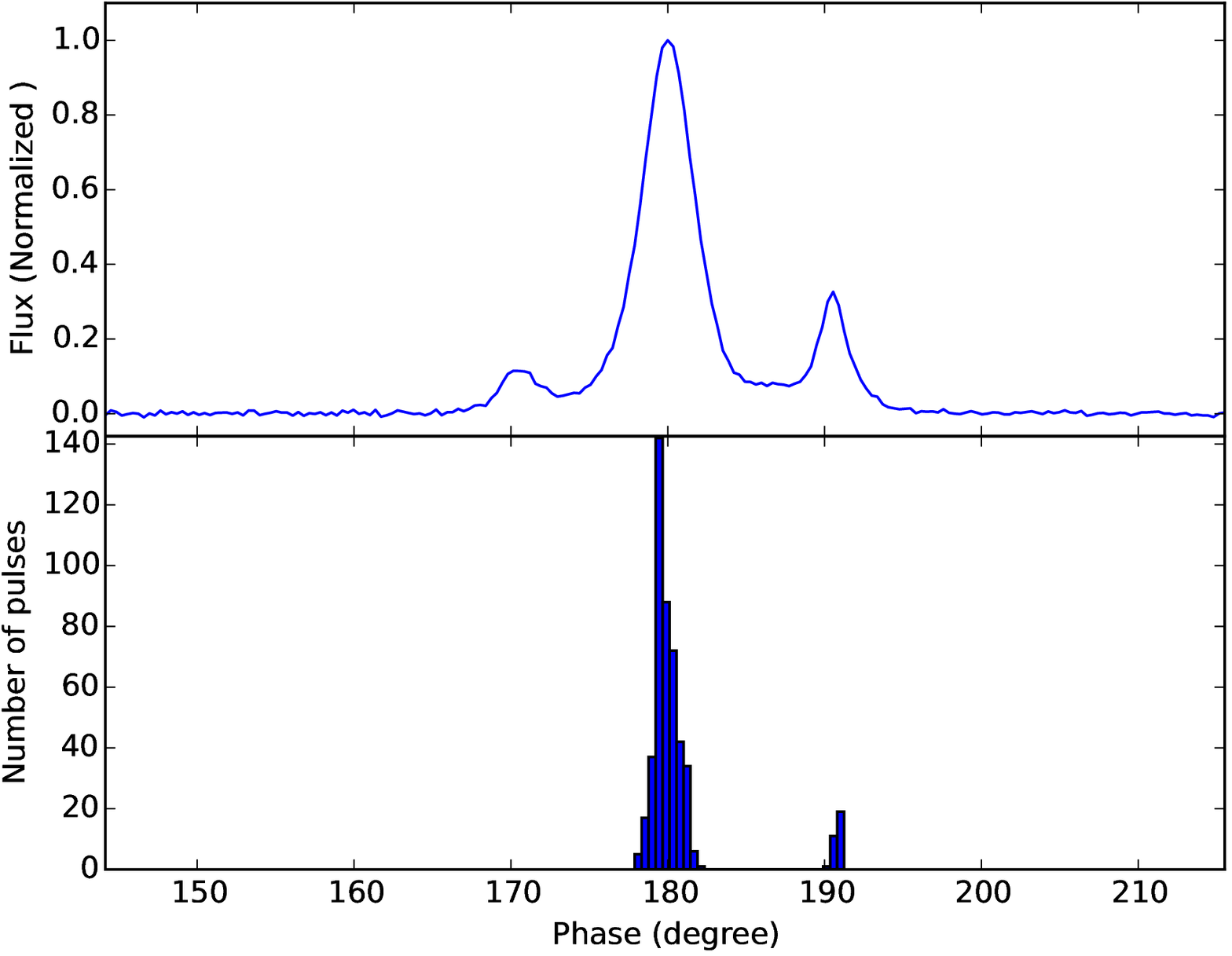}
\end{minipage}
}
\hspace{0.05\textwidth}
\subfigure[Abnormal mode]{
\begin{minipage}[b]{0.39\textwidth}
\includegraphics[width=0.88\textwidth,angle=-90]{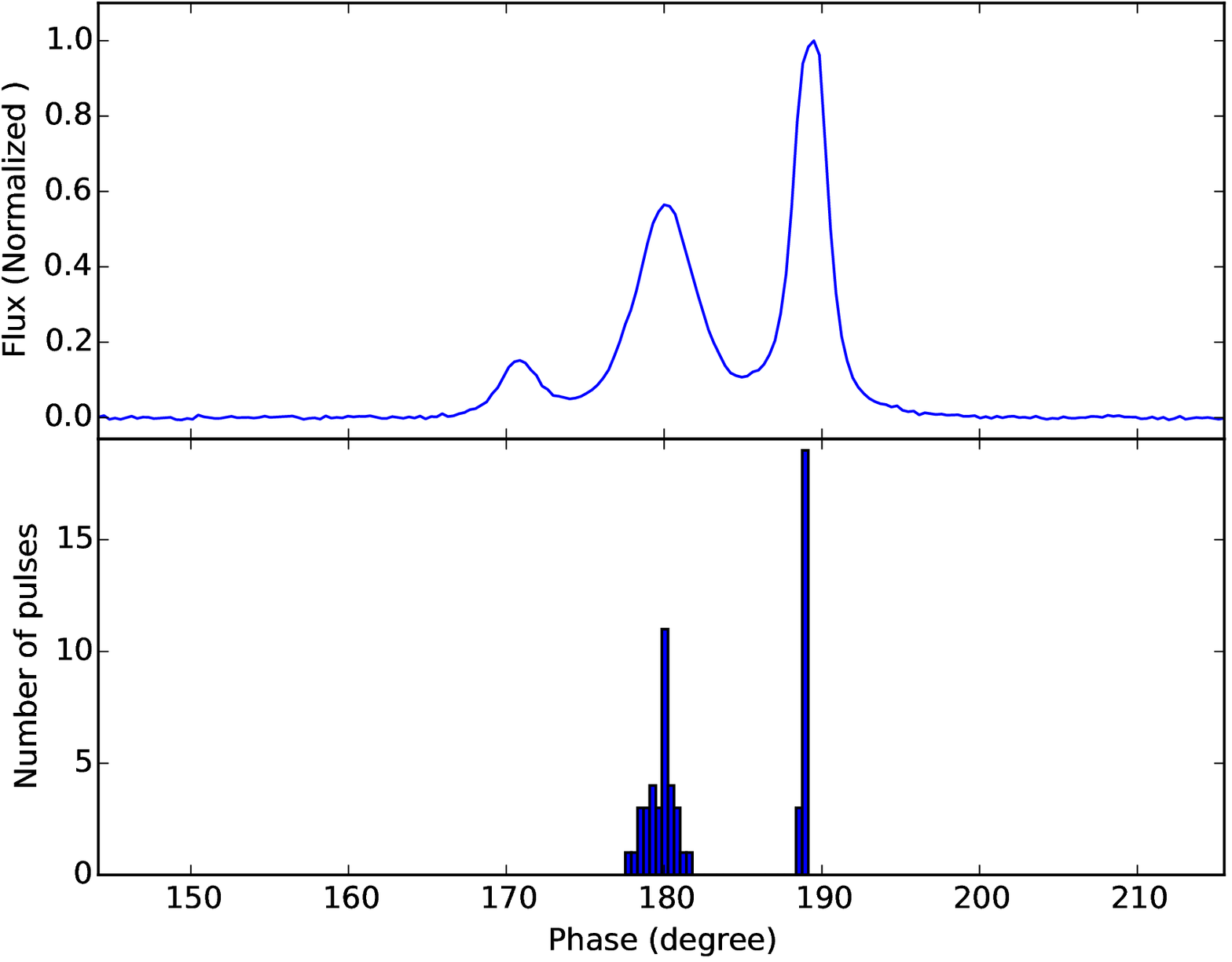}
\end{minipage}
}
\caption{Histogram of central pulse phase for PSR~B0329+54 3~cm
  band normal-mode and abnormal-mode bright pulses with $S_{\rm pk}$
  greater than 10 times $S_{\rm int,pk}$. For comparison, the
  normalized integrated pulse profile for the corresponding
  mode is shown in the top panel.}
\label{fig:hist_bright}
\end{figure*}

We also performed a statistical analysis on the phase position of bright
pulses whose $S_{\rm pk}$ was greater than 10 times $S_{\rm int,pk}$.
Data for the two bands and two modes are separately
analyzed.  In Figure~\ref{fig:hist_bright}, we present the phase
distribution of bright pulses occurring in both normal mode and
abnormal mode at 3~cm. It is clear that the bright pulses mostly
occurred at the phase of the peaks of the corresponding integrated profile.
We also detected the same behavior of the narrow bright pulses in
the 13~cm data. Finally, we performed a similar statistical analysis
on pulses with $S_{\rm pk}$ exceeding 5 times of the $S_{\rm int,pk}$ and
found the same behavior in both the normal and abnormal
modes.

\section{Analysis of subpulse drifting in PSR~B0329+54}\label{sec:drift}
Some pulsars that show mode changes also exhibit drifting
subpulses \citep{wf81, rwr05}.  Subpulse
drifting can be characterized by two parameters along orthogonal
directions: the horizontal separation between adjacent subpulses in
pulse longitude ($P_2$) and the vertical separation between
drift bands in pulse periods ($P_3$). It can be
detected using an auto-correlation analysis \citep{tmh75} or Fourier
analysis \citep{es02}. The longitude-resolved standard deviation
(LRSD) and the longitude-resolved modulation index (LRMI) are useful
diagnostics of subpulse modulations. Plots from the Fourier analysis,
such as the two-dimensional fluctuation spectrum (2DFS) and the
longitude-resolved fluctuation spectrum (LRFS), are effective tools to
characterize $P_2$ and $P_3$ \citep{wes06,wse07,wel+16}.

We first applied the auto-correlation analysis method to investigate
the subpulse drifting properties of the normal and abnormal modes at
both 13~cm and 3~cm, but found no preferred direction for subpulse
drifting. We then calculated the LRSD, LRMI, LRFS and 2DFS
mentioned above. Besides the 2DFS analysis on the entire phase range
of pulsar radiation, the 2DFS of the data corresponding to phase range
of three peaks components (C1, C2 and C3) were calculated individually
to investigate the subpule drifting in these components. We did not
attempt to further divide into smaller phase windows in the 2DFS
analysis because of the limited sensitivity. In the Fourier analysis,
the input pulse series were divided into 256-pulse blocks to improve
the sensitivity. Values from the analysis of each block were
averaged to give the final results.

\begin{sidewaysfigure}
\centering
\subfigure[Normal mode (Left:13~cm; Right: 3~cm)]{
\fbox{
\begin{minipage}[b]{0.22\textwidth}
\includegraphics[scale=0.9,angle=-90]{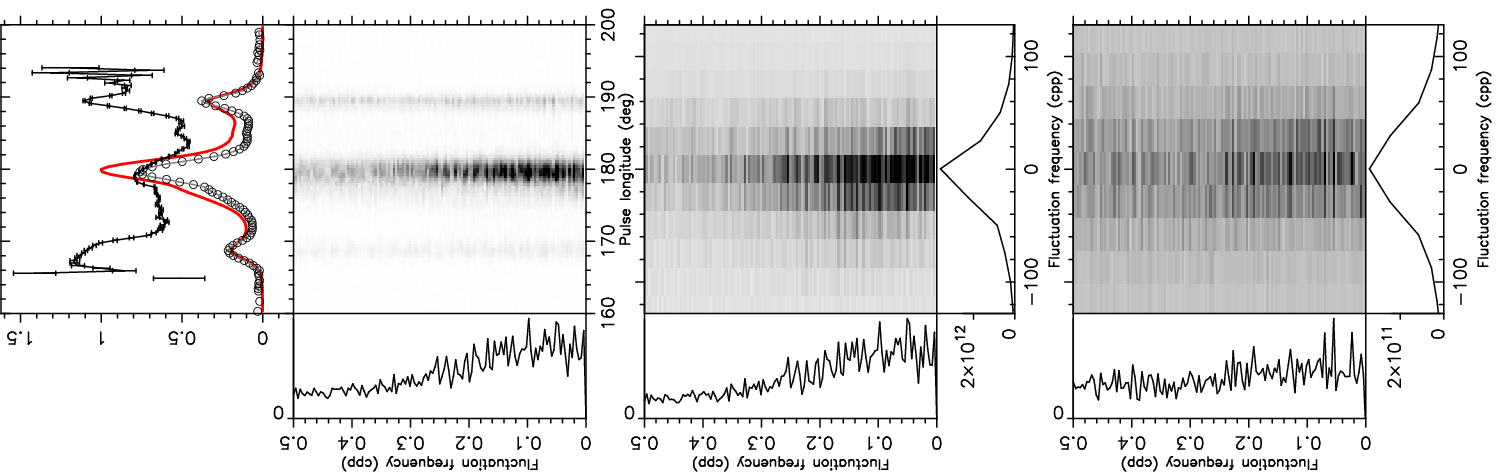}
\end{minipage}
\begin{minipage}[b]{0.22\textwidth}
\includegraphics[scale=0.9,,angle=-90]{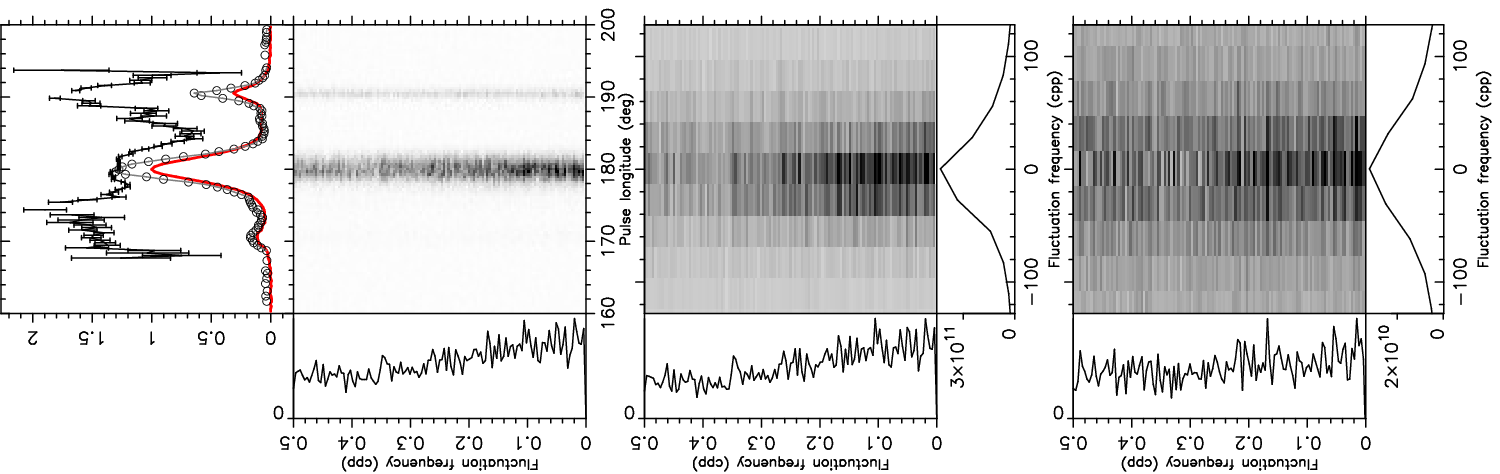}
\end{minipage}
}
}
\subfigure[Abnormal mode (Left:13~cm; Right: 3~cm)]{
\fbox{
\begin{minipage}[b]{0.22\textwidth}
\includegraphics[scale=0.9,,angle=-90]{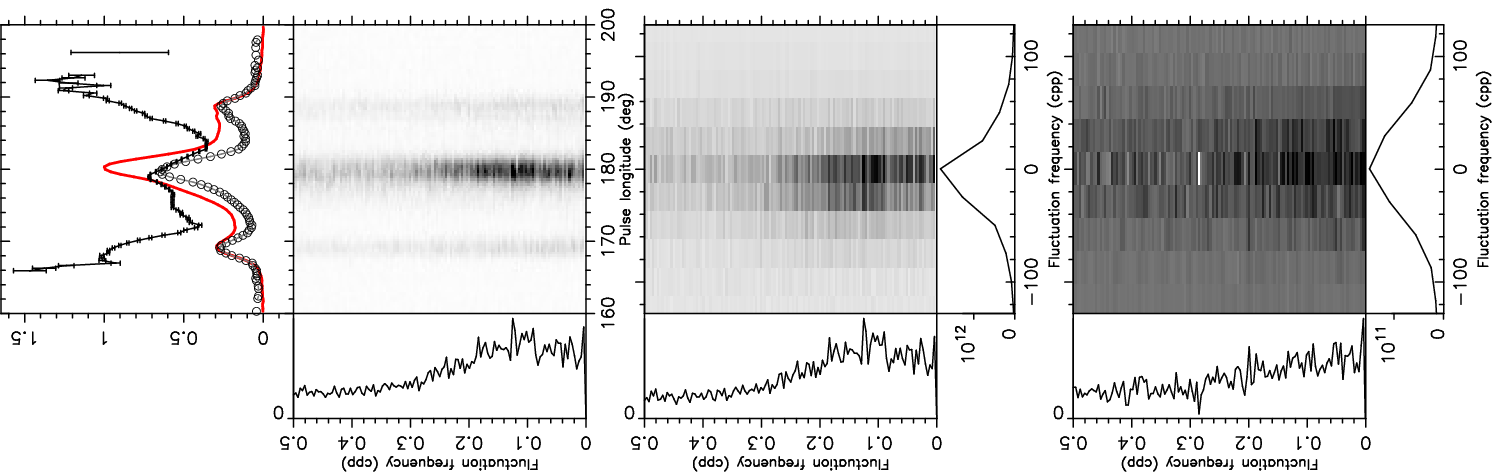}
\end{minipage}
\begin{minipage}[b]{0.22\textwidth}
\includegraphics[scale=0.9,,angle=-90]{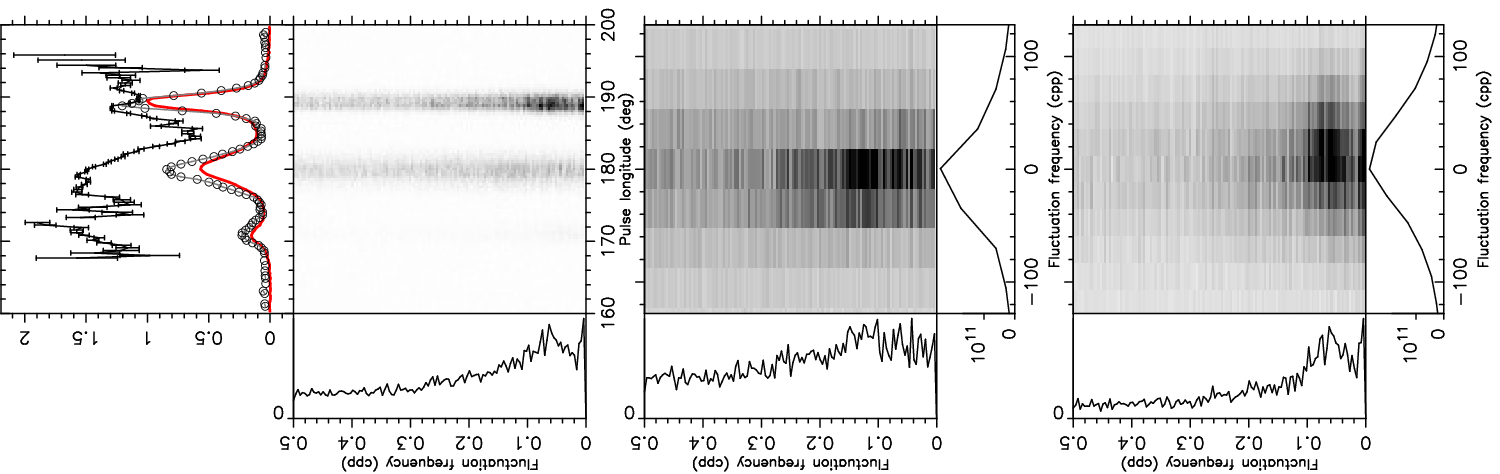}
\end{minipage}
}
}
\put(0,-40){\tiny LRSD,LRMI}
\put(0,-115){\tiny LRFS}
\put(0,-215){\tiny 2DFS-2}
\put(0,-325){\tiny 2DFS-3}

\caption{Subpulse drifting analysis results for (a) normal mode
  and (b) abnormal mode observations at both 13~cm and 3~cm for PSR~B0329+54 on
  MJD~56785.20 ($N$=16). The top row shows the longitude-resolved rms
  fluctuation and modulation index, along with the relevant integrated
  profile (in red). The second row gives the longitude resolve
  fluctuation spectrum along with the integrated spectra on the left
  side of each plot. The third and fourth rows give the
  two-dimensional fluctuation spectra for C2 and C3 respectively, with
  the integrated spectra to the left and below. }
\label{fig:lrfs}
\end{sidewaysfigure}

In Figure~\ref{fig:lrfs}, we present the results of the subpulse
analyses for the observation at MJD~56785.20 ($N$=16). In the top
panel we show the integrated pulse profiles and the LRMI and LRSD
plots. The modulation index has peaks greater than 1.0 at the phases
of peaks in the integrated profile and is generally higher toward the
wings of the profile. The LRFS in the second row shows that in the normal
mode at both bands, there is a significant broad red modulation which
is concentrated in C2. Such red modulations are commonly observed in
core components \citep{wes06, wse07}.  When the pulsar is in the
abnormal mode (right columns) the situation is quite different. At
13~cm there is still a red modulation, but unlike for the normal mode,
it has a broad peak around 0.12 cycles/period (cpp). At 3~cm, there
are more significant changes, with a prominent quasi-period modulation
at about 0.06~cpp in C3 dominating the integrated spectrum. The red
modulation can still be seen in C2, but it appears weaker compared to
the normal mode.

The lower two rows of Figure~\ref{fig:lrfs} show 2DFS plots for C2 and
C3. (The LRFS plots show no significant modulation in C1.) The
horizontally averaged modulation for C2 at 3~cm shows that the broad
quasi-periodic modulation around 0.12~cpp seen in C2 at 13~cm is also
present in this component at 3~cm. Although there are some asymmetries
in the vertically averaged modulation spectra, they are all centered
on zero frequency within the uncertainties. There is therefore no
clear evidence for subpulse drifting in this pulsar. However, because
of the observed quasi-periodic features, the pulsar can be labelled as
type ``Dif*'' according to the classification of \citep{wes06}.

\section{Pulse nulling analysis}\label{sec:null}
\begin{figure}[ht]
\centering
\includegraphics[width=0.5\textwidth,angle=-90]{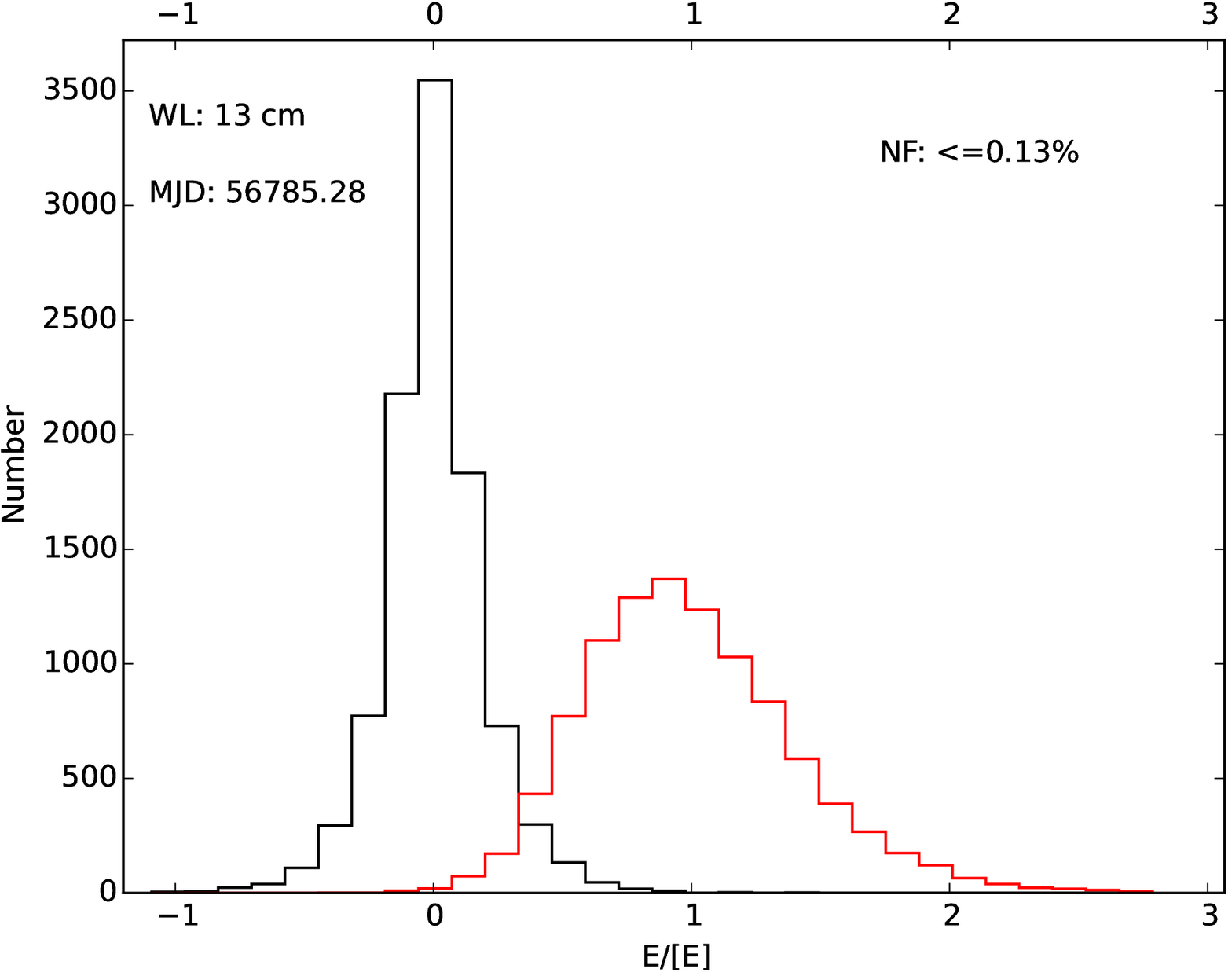}
\par\vspace{14pt}
\includegraphics[width=0.5\textwidth,angle=-90]{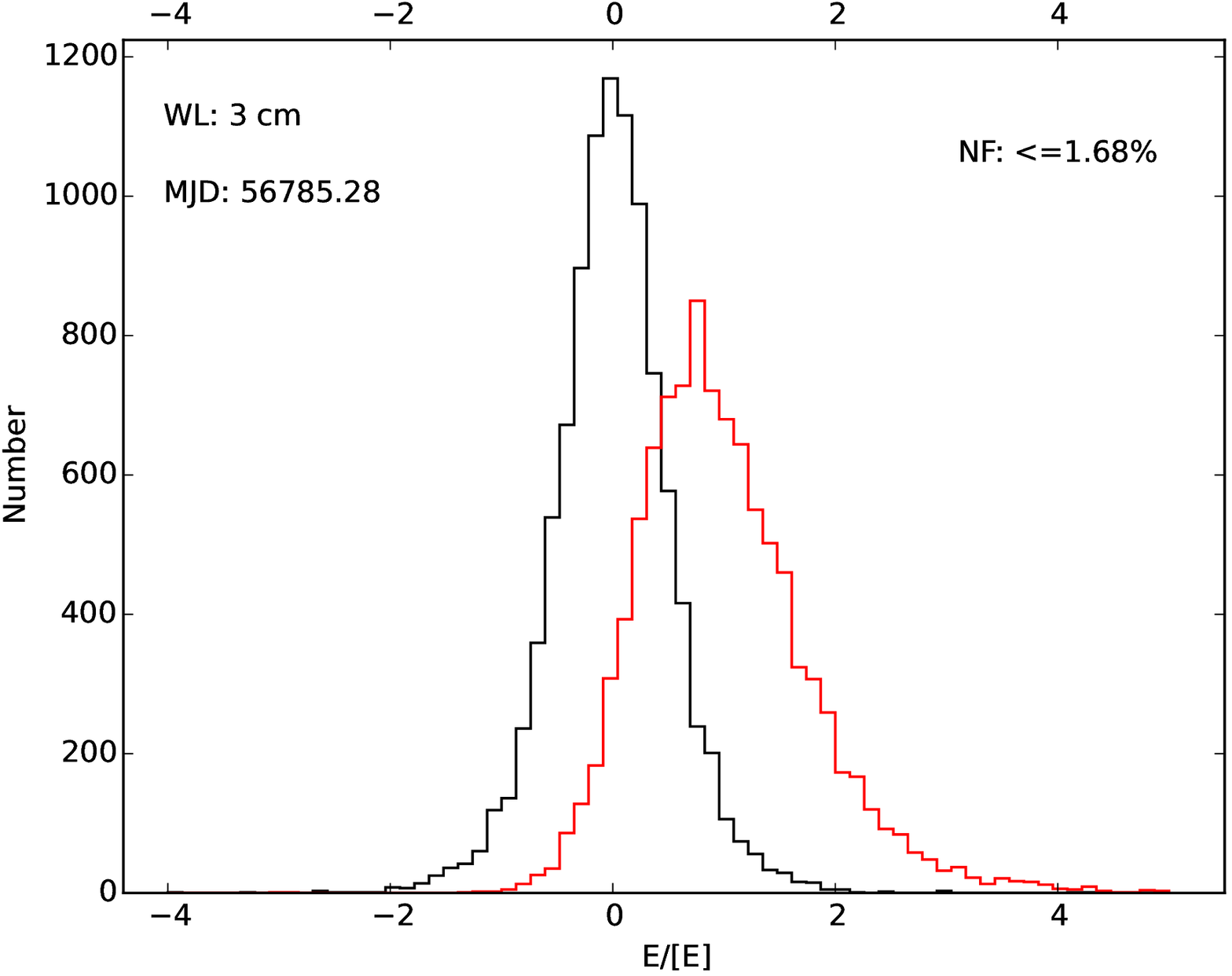}
\caption{On-pulse (red) and off-pulse (black) pulse-energy histograms
  for PSR~B0329+54 in the 13~cm (upper panel) and 3~cm bands (lower
  panel) from the observation at MJD 56785.28.}
\label{fig:nulling}
\end{figure}

Previous studies indicated that pulse nulling tends to be more common
in the population of pulsars with multi-component integrated profiles,
and these pulsars generally have larger characteristic ages
\citep{ran86,big92a}. It was also suggested that pulse nulling could
be a manifestation of mode change, for which one mode has extremely
weak radiation \citep{elg+05,wmj07,syh+15}. A challenge for nulling analyses
is the often limited signal-to-noise ratio of single-pulse
observations. In our analysis we used the method in Smith~(1973) and
Ritchings~(1976)\nocite{smi73,rit76}. A pulsar OFF window with the
same width as the ON window is selected to quantify the noise
properties of the signal. To reduce the effects of interstellar
scintillation, a normalized energy is obtained for 500-pulse blocks by
dividing by the mean energy of each block. The OFF data in the
corresponding block are also divided by the same factor and histograms
of the normalized ON and OFF pulsar data are computed. ON and OFF
histograms for 13~cm and 3~cm for the observation on MJD~56785.28
($N$=17), which has combatively high signal-to-noise ratio, are
presented in Figure~\ref{fig:nulling}. This figure shows that there was
no significant population of null pulses. The procedure for obtaining
an upper limit on the null fraction is as follows. An increasing fraction of the
OFF probability distribution is subtracted from the ON probability
distribution until some portion of the residual distribution becomes
either negative or discontinuous. Using this method, we found that the
upper limits on the null fraction for PSR~B0329+54 are 0.13\% and 1.68\%
at 13~cm and 3~cm, respectively. This result is consistent with the
0.25\% limit obtained at 408~MHz by \citet{rit76}.

\section{Discussion and conclusions}
\label{sec:discuss}
In this study, we obtained 41.6 hours of simultaneous 13~cm and 3~cm
single-pulse data for PSR~B0329+54 over several epochs with the
TMRT. We found two distinct modes of the radio emission, referred to as normal
mode and abnormal mode respectively, with the pulsar in the
normal mode 87\% of the time. The mode change phenomenon was more obvious
at 3~cm than at 13~cm, with the third principal component of the
integrated profile, C3, shifted to an earlier phase
and substantially stronger in the abnormal mode. On the
other hand, at 13~cm, C3 was weaker and became part of a flat ``shoulder''
connecting to C2 in the abnormal mode. Our results show that the mode
switch occurred synchronously at 13 and 3~cm within a few rotation
periods. This result is in agreement with previous observations
\citep{bmsh82}.

The shape of the integrated profile of pulsars gives important
  information about the structure of emission regions in pulsar
  magnetospheres.  In the polar-cap model \citep{gj69,rs75}, the
  observed radio pulses result from coherent radiation of relativistic
  particles flowing along dipolar field lines emitted in the
  tangential direction.  The polar cap on the neutron star surface
  is defined by the last open magnetic field lines. In this dipolar
  model, assuming a symmetrically illuminated beam, the opening semi-angle
  $\rho$ of the pulsar beam corresponding to an overall pulse width $W$
  is given by:
\begin{equation}
\rho=2\arcsin[\sin(\alpha+\beta) \sin\alpha \sin^2(W/4) + \sin^2(\beta/2)]^{1/2},
\end{equation}
where $\alpha$ is the angle between the rotation and magnetic axes and
$\beta$ is the angle of closest approach of the line of sight to the magnetic
axis \citep{ggr84}. The
radio emission altitude, which is the radial distance of the radio emission
region measured from the center of the star, can be approximated for
small opening angles by:
\begin{equation}
r\simeq R_{\rm NS} P (\rho/1.24\degr)^2,
\end{equation}
where $R_{\rm NS}$ is the neutron-star radius, $P$ is the pulsar
period in seconds and $\rho$ is the emission cone half-opening angle
in degrees \citep{gk93}.

In Table~\ref{tab:intprof} we give estimates of $\rho$ and  $r$ for
PSR~B0329+54 for the two modes and two bands based on these
relations. We take $\alpha = 30.4\pm0.4 \degr$ and $\beta = 2.5\pm0.4\degr$
\citep{lm88,ran93b} and use values of $W_{10}$ from
Table~\ref{tab:intprof}. Since the observed pulse widths are somewhat
smaller at 3~cm, the derived values of $\rho$ and $r$ are also smaller
for emission in this band. There is little or no significant
dependence of $\rho$ or $r$ on the emission mode.

\begin{figure}[ht]
\centering
\includegraphics[width=0.5\textwidth,angle=-90]{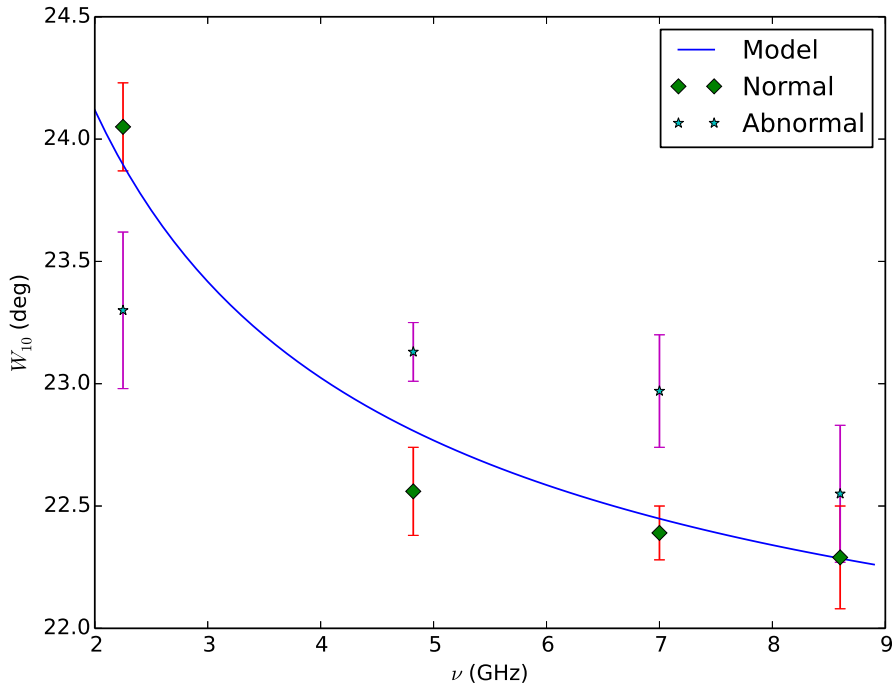}
\caption{The 10\% peak pulse width $W_{\rm 10}$ vs the observation frequency $\nu$. The $W_{\rm 10}$
results at 4.8~GHz and 7.0~GHz were also obtained using the TMRT.}
\label{fig:w10freq}
\end{figure}

The frequency dependence of the pulse width ($W_{\rm 10}$) was
  studied previously for a large sample of normal pulsars including
  PSR~B0329+54 \citep{cw14}. It was found that the observed $W_{\rm 10}$
  could be well fitted with the relationship:
\begin{equation}
W_{\rm 10}=A\nu^{\mu}+W_{\rm min},
\end{equation}
where the widths are in degree and the frequency $\nu$ is in GHz.
For PSR~B0329+54, the best-fit coefficient $A$, index $\mu$ and
asymptotic constant $W_{\rm min}$ is 4.67, -0.68 and $21.2\degr$,
respectively. Figure~\ref{fig:w10freq} shows this
model fit along with the observed $W_{\rm 10}$ values at 13~cm
(2.3~GHz) and 3~cm (8.6~GHz) from Table~\ref{tab:intprof} along with
other TMRT measurements at 4.8~GHz and 7.0~GHz \citep{yan18}  It can be
seen that our results for the normal mode follow the relationship very
well as expected since the \citet{cw14} results would have been
dominated by normal-mode emission. However, for the abnormal mode, the
frequency dependence is much flatter, with the observed widths
narrower at 2.3~GHz, but wider for $\nu > 5$~GHz.

As the last two columns of Table~\ref{tab:obsinfo} show, the
  flux densities of the abnormal and normal modes show different
  fluctuation properties across epochs at 13~cm and 3~cm. The
  parameter $S_{\rm A/N,13cm}$ was randomly distributed around 0.0,
  while $S_{\rm A/N,3cm}$ was generally larger than 0.0. We believe
that the fluctuation of $S_{\rm A/N,13cm}$ was caused by the
interstellar scintillation. Based on previous interstellar
scintillation results on this pulsar at 1540~MHz \citep{wymw08}, it
can be inferred that the de-correlation bandwidth and typical
scintillation time-scale at 13~cm are about 50~MHz and 30~min
respectively, assuming a Kolmogorov fluctuation spectrum for
interstellar medium. In our case, the 13~cm observations have a
bandwidth of 100~MHz and tens of minutes duration and so only a few
`scintiles' were included in the estimate of $S_{\rm A/N,13cm}$. Hence
the effects of interstellar scintillation on the flux
density measurements were substantial.

On the other hand, the modulation of $S_{\rm A/N,3cm}$ cannot be
simply interpreted as the interstellar scintillation effect. The
interstellar scintillation is much weaker at 3~cm
\citep{mss+96}. Previous low frequency diffractive interstellar
scintillation observation results on this pulsar \citep{wymw08} and
related theory \citep{ric90} suggest a timescale of 1.6~hr for the
weak interstellar scintillation. Although the scintillation effects
were still evident, if scintillation were the main cause of the
variations, the average ratio would be around 0.0. In fact,
$S_{\rm A/N,3cm}$ were almost always greater than 0.0. This reflects an
intrinsic flux density variation between the normal and abnormal
pulsar modes, largely due to the flux enhancement of C3 in the
abnormal mode at 3~cm.

Previous studies found possible correlations between mode changes
and subpulse drifting. Our multi-epoch observations indicated that, in
the abnormal mode, PSR~B0329+54 showed evidence for quasi-periodic
fluctuations. For C2 there was a broad spectral feature centered
around 0.12~cpp, corresponding to a $P_3$ in the range of 5 -- 20
periods. For C3, no quasi-period modulation was evident, but at 3~cm
there was a clear feature centered at 0.06~cpp, that is, $P_3\sim 15$
pulse periods. This feature was not seen in previous studies at
21~cm and 92~cm \citep{wes06, wse07}. Although the 2D fluctuation
spectra had some asymmetry when integrated vertically, in general
there was no significant offset from zero, and hence no evidence for
pulse drifting. Previous low frequency observations are also
consistent with no preferred drifting directions for this pulsar.

We found no evidence for pulse nulling in PSR~B0329+54 with upper
limits on the null fraction of 0.13\% and 1.68\% at 13~cm and 3~cm,
respectively. The higher limit at 3~cm is principally because of the
lower S/N of these observations. Previous low frequency (408~MHz)
observation gave a consistent limit on the null fraction of 0.25\%
\citep{rit76}. It is clear that PSR~B0329+54 is a pulsar with very low
probability of nulling. Previous investigations indicated that pulsar null
fractions correlate more with the characteristic age than with the pulse
period \citep{wmj07}. Given that the characteristic age of
PSR~B0329+54 is 5.53~Myr \citep{hlk+04}, a low null fraction is not
surprising.

Single-pulse analyses show the existence of occasional narrow bright
pulses that seem to be a statistically separate population compared to
the distribution of other single
pulses. Figure~\ref{fig:widthpeak} shows that strong single
pulses tend to be narrower than the average width, although there is
no clear correlation between width and strength. Though the peak
amplitudes of these bright pulses are relatively high, their total power
is not as strong as the giant pulses detected in a few pulsars such as
the Crab pulsar (Knight 2006 and references therein)\nocite{kni06}.
In PSR~B0329+54, the bright pulses preferentially occur at the pulse
phases of the second and third peaks of the integrated profile.
The burst rate at 3~cm is up to 100 times higher than that at 13~cm.
This suggests a real difference in the coherence properties of the
emission process for observed frequencies around 10~GHz compared to
lower frequencies, possibly related to an emission location closer
to the neutron-star surface. Future studies with simultaneous
multi-frequency observations of a large pulsar sample will be important
in helping to understand these differences.

\section*{Acknowledgments}
This work was supported in part by the National Natural Science
Foundation of China (grants U1631122, 11403073 and 11633007), National
Basic Research Program of China (973 program) No.~2012CB821806, the Strategic
Priority Research Program ``The Emergence of Cosmological Structures"
of the Chinese Academy of Sciences, Grant No.~XDB09000000 and the
Knowledge Innovation Program of the Chinese Academy of Sciences (Grant
No.~KJCX1-YW-18) and the Scientific Program of Shanghai Municipality
(08DZ1160100). We would like to thank the anonymous referee
for helpful suggestions.

\label{lastpage}


\end{document}